\documentclass[twoside]{article}

\usepackage{float}
\usepackage{color}
\usepackage{greekbf}
\usepackage{float}
\usepackage{amsmath,amssymb}
\usepackage{mathrsfs}
\usepackage{enumerate}
\usepackage{srcltx}
\usepackage{mathtools}
    \mathtoolsset{showonlyrefs}
\usepackage{graphicx, psfrag}
\usepackage{setspace}
\usepackage{hyperref}
\usepackage[numbers]{natbib}
\usepackage{subfig}
\hypersetup{bookmarks=false,
colorlinks=false,
linkcolor=black,
urlcolor=black,
pagebackref=true,
pdfstartview={FitH},
bookmarksopenlevel=2
}

\newtheorem{remark}{Remark}

\DeclareMathAlphabet{\mathbf}{OT1}{cmr}{bx}{it}
\definecolor{red}{rgb}{0.9,0,0}
\definecolor{blue}{rgb}{0.2,0.2,0.8}
\definecolor{green}{rgb}{0.0,0.5,0.2}
\definecolor{darkblue}{rgb}{0.2,0.2,0.5}
\definecolor{orange}{rgb}{1,0.5,0}
\definecolor{pink}{rgb}{0.96,0.5,0.46}
\definecolor{lblue}{rgb}{0.18,0.74,1}
\definecolor{cyan}{rgb}{0,0.8,0.8}

\newcommand {\Cc}  {\mathcal{C}}

\newcommand {\Ec}  {\mathcal{E}}

\newcommand {\Ao} {\mathbb{A}}

\newcommand {\Do} {\mathbb{D}}

\newcommand {\Fes} {\mathcal{F}}

\newcommand {\Bb} {\mathbf{B}}
\newcommand {\Cb} {\mathbf{C}}

\newcommand {\Eb} {\mathbf{E}}

\newcommand {\Kb} {\mathbf{K}}

\newcommand {\Ib} {\mathbf{I}}

\newcommand {\Nb} {\mathbf{N}}

\newcommand{\tR}[1]{{\color{black} #1}}
\newcommand{\ov}{\overline}

\DeclareMathOperator{\dd}{\textrm{d}\!}
\DeclareMathOperator{\curl}{\textrm{curl}}
\begin{document}
	
\doublespacing

\title{\vspace{-3cm} {\bf  Enhanced models for the nonlinear bending of planar rods:
		localization phenomena and multistability}}

\author{
Matteo Brunetti$^1$\!\!\!\!\! \and  Antonino Favata$^2$\!\!\!\!\! \and Stefano Vidoli$^{3}$
}

%
%\date{\today}

%%% ----------------------------------------------------------------------
\maketitle
%%% ----------------------------------------------------------------------

\vspace{-1cm}
\begin{center}
	{\small
		$^1$ Department of Civil and Industrial Engineering\\ University of Pisa, Pisa, Italy\\
		\href{mailto:matteo.brunetti@unipi.it}{matteo.brunetti@unipi.it}\\[8pt]
		$^2$ Department of Structural and Geotechnical Engineering\\
		Sapienza University of Rome, Rome, Italy\\
		\href{mailto:antonino.favata@uniroma1.it}{antonino.favata@uniroma1.it}\\[8pt]

		$^3$ Department of Structural and Geotechnical Engineering\\
		Sapienza University of Rome, Rome, Italy\\
		\href{mailto:stefano.vidoli@uniroma1.it}{stefano.vidoli@uniroma1}\\[8pt]
	}
\end{center}

\pagestyle{myheadings}
\markboth{M.~Brunetti, A.~Favata, S.~Vidoli }
{Enhanced one-dimensional rods}

\vspace{-0.5cm}
\section*{Abstract}
We deduce a 1D model of elastic planar rods starting from the F\"{o}ppl--von K\'{a}rm\'{a}n model of thin shells. Such model is enhanced by additional kinematical descriptors that keep explicit track of the compatibility condition requested in the 2D  parent continuum, that in the standard rods models are identically satisfied after the dimensional reduction. An inextensible model is also proposed, starting from the nonlinear Koiter model of inextensible shells.
These enhanced models describe the nonlinear planar bending of rods and  allow to account for some phenomena of preeminent importance even in 1D bodies, such as formation of singularities and localization (d-cones), otherwise inaccessible by the classical 1D models. Moreover, the effects of the compatibility translate into the possibility to obtain multiple stable equilibrium configurations.

\vspace{1cm}

\noindent {\bf Keywords}: dimensional reduction, rod theory, d-cone, localization, multistability

\tableofcontents

\vspace{.5cm}

%\noindent \large{\textbf{Graphical Abstract}}
\normalsize
 
%\vspace{.2cm}
% 
%\begin{figure}[h!]
%	\begin{center}
%		\includegraphics[scale=.22]{GraphicalAbstract.pdf}
%	\end{center}
%\end{figure}

\newpage
\section{Introduction}
In this paper we intend to propose two mathematical models of a class of bodies that are \textit{thin} and \textit{slender} at the same time, a feature that allows to have recourse to a  1D  theory. The moderately large deflection of thin elastic plates or shells, \textit{i.e.}, bodies which are intrinsically \textit{thin}, can be well described by the F\"{o}ppl--von K\'{a}rm\'{a}n (FvK)  model \cite{Foppl,vK}\cite{Ciarlet,Ciarlet1980}. Because of the smallness parameter given by the thickness, such a model is intrinsically 2D.  

On the  side of \textit{slender} bodies, besides the classical  rod models, in the literature are often adopted a  number of refinements, many of them having the scope to go beyond the limits of the standard rod models, such as Timoshenko's or Euler's. 

\noindent A full description of the huge literature on the subject is unattainable; we here quote \cite{Cimetiere1988,Coleman1988,Sanchez1999,Hamdouni2006} and, among the most recent works \cite{guinot,Audoly2016,Lestringant2017,Geymonat2018,ballard_2018,Lestringant2018,Lestringant2019,Calladine2019}. In particular, in \cite{ballard_2018} a model for  rods and thin-walled rods  is rigorously obtained from a formal asymptotic analysis of three-dimensional linear elasticity. In \cite{Lestringant2019}  a general  method for deriving one-dimensional models for nonlinear structures has been proposed; the models capture the contribution to the strain energy arising not only from the macroscopic elastic strain, but also from the strain gradient.  Moreover, the
so-called models \`{a} la \tR{Sadowsky}, usually generated starting from plate models, are sometimes adopted. Among these, the original one proposed by Sadowsky in 1930 \cite{sadowsky_1930} and formally justified by Wunderlich in 1962 \cite{wunderlich_1962}, has been deduced from the linear Kirchhoff plate model. Recently, similar models have been deduced from the non-linear von K\'{a}rm\'{a}n plate model \cite{freddi_2016_1,freddi_2018}; the limit problems penalize extensional, flexural and torsional deformation and they are comparable to classical non-linear rod theories. 

In \cite{freddi_2012,freddi_2013} a hierarchy of one-dimensional models for thin-walled rods with rectangular cross-section, starting from three-dimensional nonlinear elasticity has been deduced. The different limit models are distinguished by the different scaling of the elastic energy and of the ratio between the sides of the cross-section. 

In \cite{Davini_2014,Davini_2014_1} the authors consider a rod whose cross section is a tubular neighborhood, with thickness scaling with a parameter $\delta_\varepsilon$, of a simple curve  whose length scales with $\varepsilon$; to model a thin-walled rods they assume that $\delta_\varepsilon$ goes to zero faster than $\varepsilon$, and they measure the rate of convergence by a slenderness parameter. The approach recovers in a systematic way, and gives account of, many features of the rod models in the theory of Vlasov.

In this paper, we  deduce two 1D models of elastic rods enhanced by additional kinematical descriptors that keep explicit track of the \textit{compatibility condition} requested in the 2D  parent continua; in the classical models this condition is identically satisfied after the dimensional reduction. The models differ for the possibility to account or not for extensibility. They  allow to describe some phenomena of preeminent importance \tR{in non-linear elasto-statics}, such as formation of singularities and \textit{localization} of the elastic energy (d-cones, elastic folds, etc.), otherwise inaccessible by the classical 1D models. Indeed, these phenomena are expression of a complex interaction between elasticity and geometry having an intrinsically 2D character, the compatibility conditions being the formal expression of such interaction. In the FvK model, e.g., the compatibility condition descends from the Gauss Theorema Egregium and expresses the relation between membrane deformations and variation of Gaussian curvature and, on selecting the isometries, identifies those changes of configuration that are energetically favorable. Moreover, the 1D compatibility condition, by  introducing a strong non-linearity in the problem, induces the possibility to have \textit{multiple stable solutions}, in accordance with experimental evidence \cite{brunetti2018,brunetti2018a}.

The paper is organized as follows. In Sec. \ref{problemsetup}, the prototypical problem we intend to face is described in detail. In Sec. \ref{sec:inex} we present a dimensional reduction to obtain a rod model starting from a 2D inextensible shell model; the problem translates into a constrained minimization, in terms of two kinematical decriptors, \textit{i.e.}, the axial and the transversal curvatures. The compatibility condition is nothing but a suitable version of the inextensibility constraint.

In Sec. \ref{sec:ex} we present a dimensional reduction starting from the FvK model. We obtain a non-local model, governed by three scalar fields:  the axial and the transversal curvatures, and the 1D counterpart of the stress Airy function. As it happens in the FvK model, these fields are not independent and the 1D compatibility prescribes how they have to be related. 

Sec. \ref{results} is devoted to results. Solving the inextensible problem translates into a simple geometric equivalent construction, that allows to obtain analytical results:  the constrained energy minimization problem is reduced to find a sequence a points on a three-dimensional cone, having minimal total distance from a given point, representing the stress-free configuration. Analytical solutions are not possible for the extensible case, and we then use a finite element method to solve the problem. We then present a comparison between the results obtained with the two enhanced rod models here formulated and the FvK predictions, in terms of displacements and stresses.

Discussions and conclusions are in Sec. \ref{discuss}. In particular, we  discuss the role of the 1D compatibility condition, showing that it is crucial to capture multistability: besides the configuration with a localized axial curvature, a second configuration is predicted, in which the transversal curvature is null and the axial one is constant. This is in agreement with   numerical results obtained with the FvK model \tR{and experimental evidence already discussed and published by some of the authors \cite{brunetti2018a}.}

%
%
%We confine the attention to a class of bodies having   two characteristic  dimensions   much  smaller  than  the length, and then efficiently modelled as a one-dimensional continuum. Our purpose to to introduce some  kinematical descriptors richer than the usual ones, such that an explicit track of the compatibility constraint \eqref{vKcomp} is retained, a feature that in classical models is not achievable. 

\subsection{Problem set-up}\label{problemsetup}
Although our theory is relevant in many circumstances, we confine the attention to a specific \tR{prototypical problem having several applications in the design
	of morphing structures, such as turbine blades or air inlets \cite{Daynes2009,Mattioni2009,Panesar2012}.}

\begin{figure}
	\begin{center}
		\includegraphics[width=0.5\linewidth]{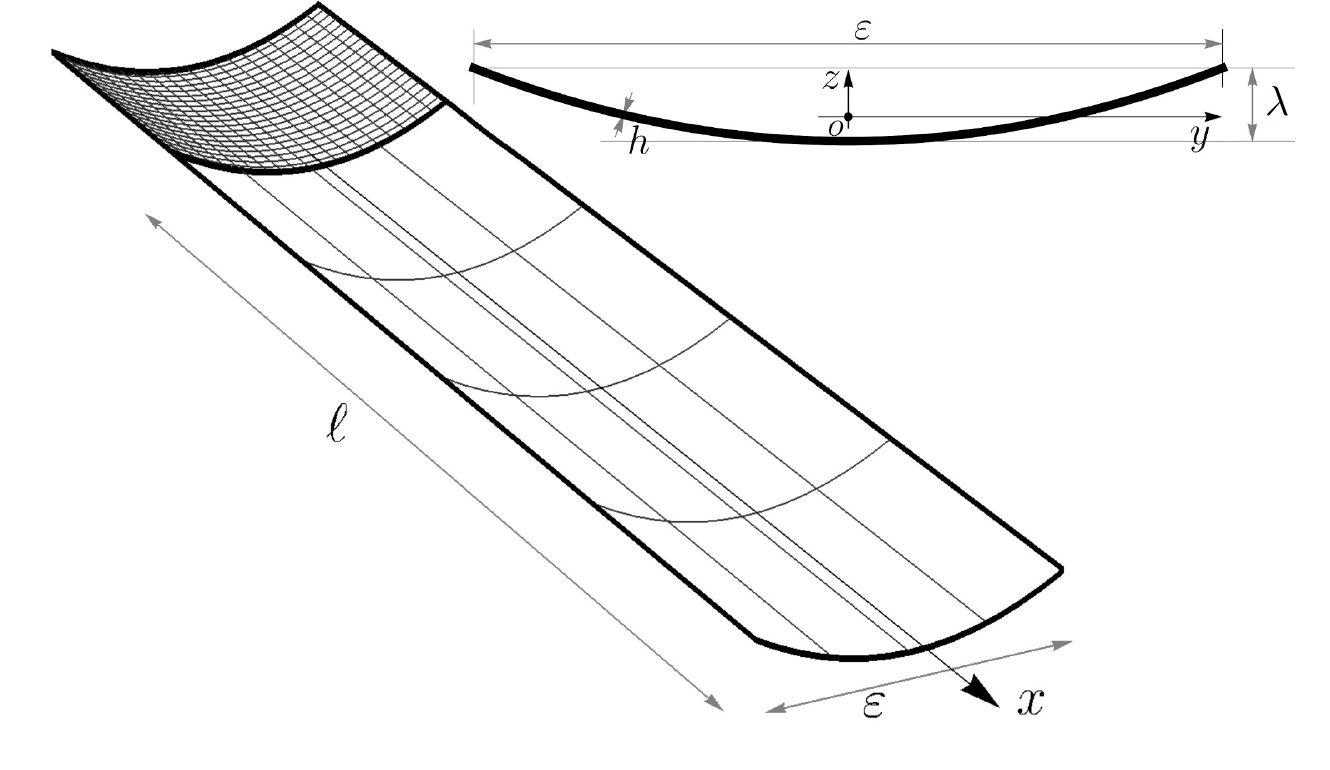}
	\end{center}
	\caption{Cylindrical stress-free configuration of the considered shell. After clamping the gray part is constrained to become flat. The main geometric quantities are shown; in particular $\ell$ indicates the effective length from the clamped side to the free-end.}
	\label{fig:geometry}
\end{figure}

\begin{figure}
	\begin{center}
		\small
		%	\hfill
		\subfloat[][]{
			\includegraphics[width=0.5\linewidth]{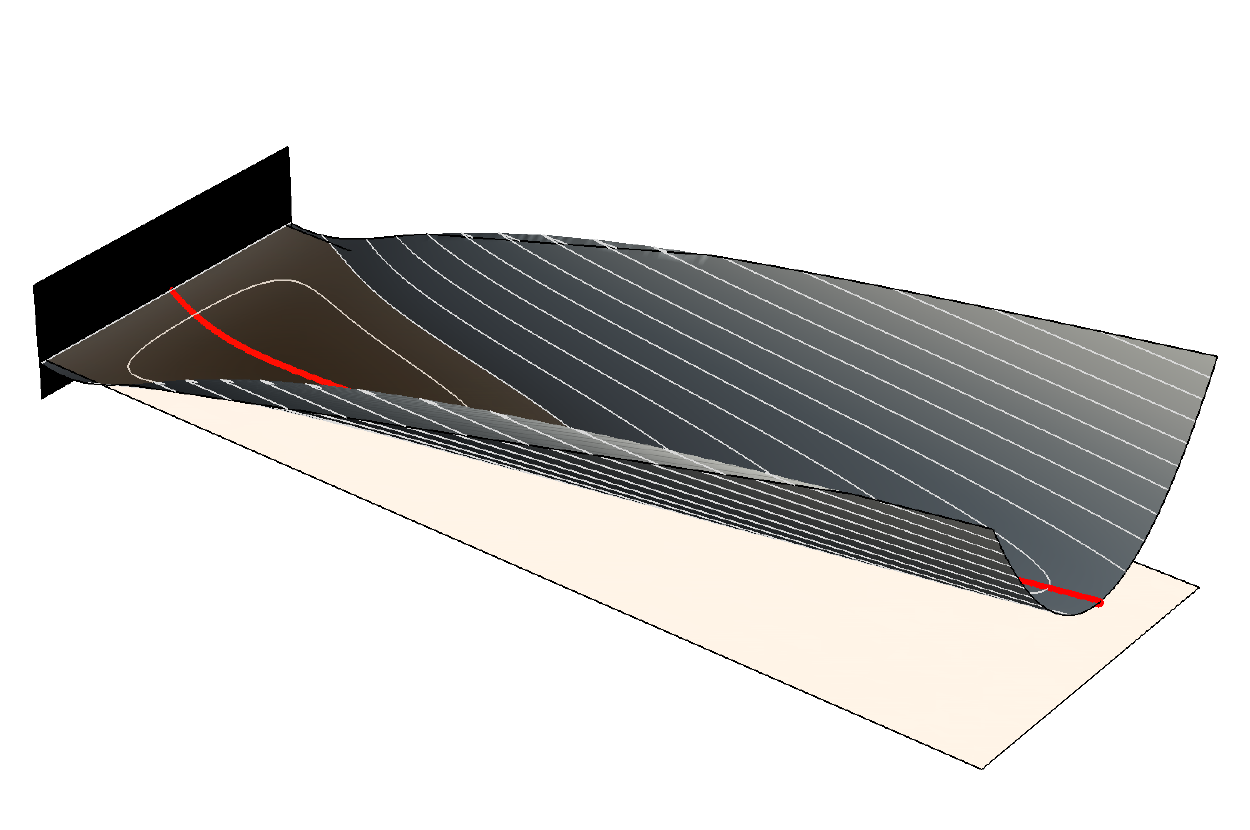}
		}
		%	\hfill
		\subfloat[][]{
			\includegraphics[width=0.45\linewidth]{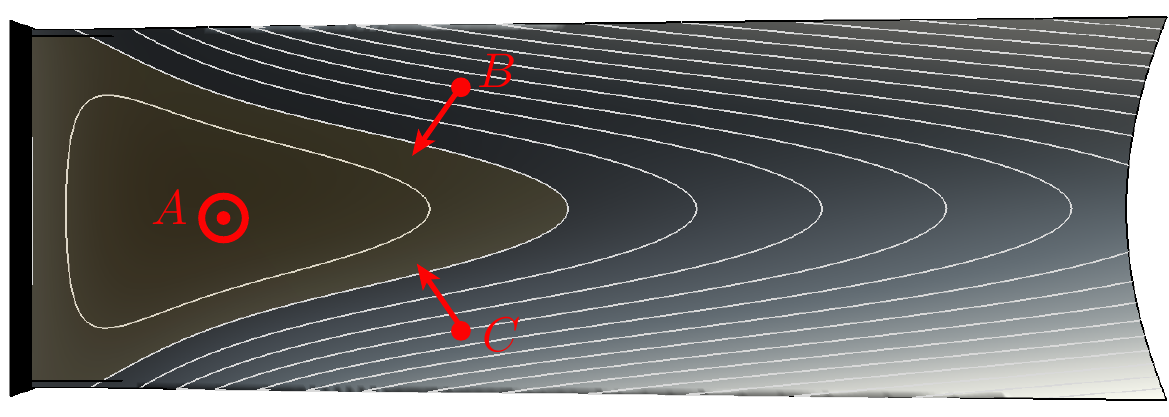}
		}
		
		\subfloat[][]{
			\includegraphics[width=0.5\linewidth]{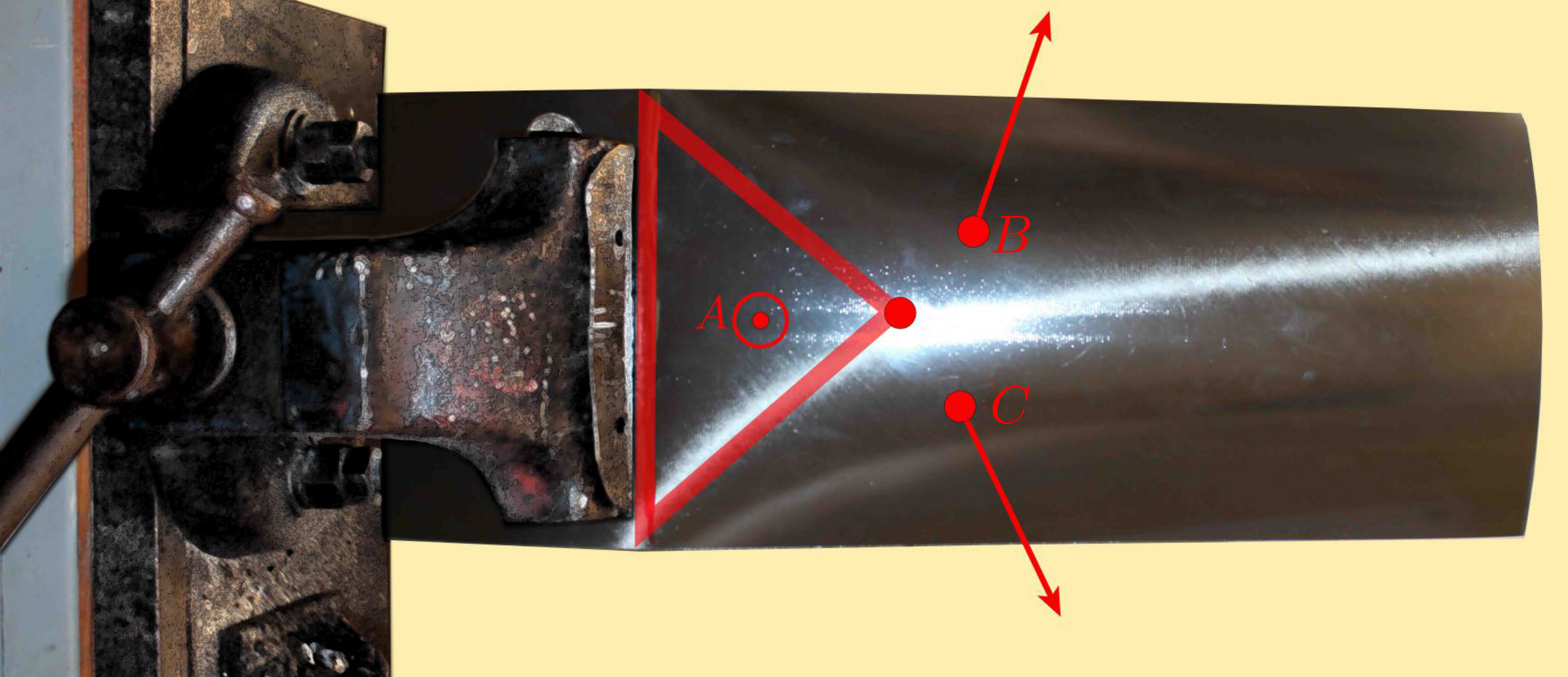}
		}
		\subfloat[][]{
			\includegraphics[width=0.5\linewidth]{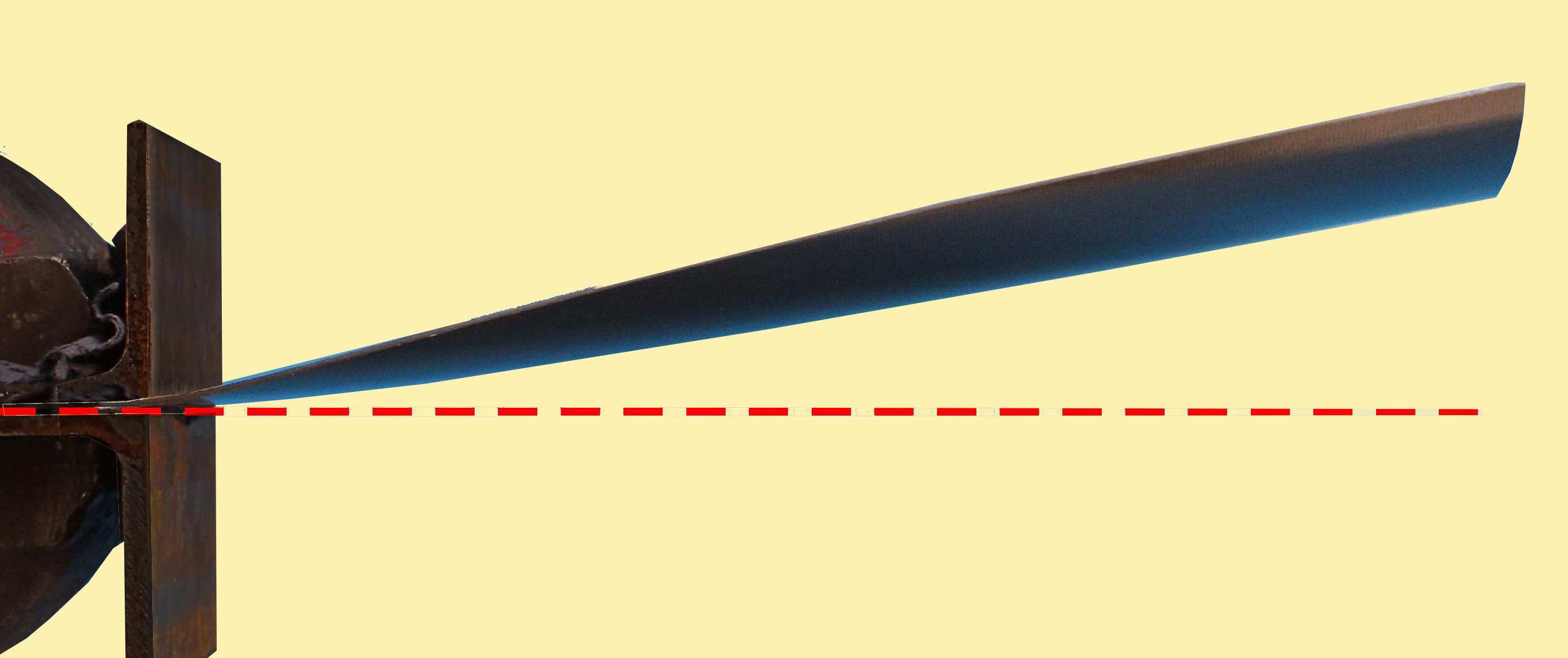}
		}
		\hfill
		\caption{Clamping a thin cylindrical shell: configuration with localized curvature, axonometric view (a), top view (b)\tR{, down view (experiment) (c),  lateral view (experiment) (d)}. The variation of Gaussian curvature localizes within the region near the clamp. The red arrows sketch the normals to the shell surface in the points $A$, $B$ and $C$.}
		\label{fig:experiment}
	\end{center}
\end{figure}

We consider a shell which in its initial stress-free natural configuration is cylindrical, shallow and has a rectangular planform, see Fig.~\ref{fig:geometry}. A portion of this shell, the one indicated by gray pattern in Fig.~\ref{fig:geometry}, is constrained to become flat after the application of a suitable clamp \cite{brunetti2016}. Clearly, this clamping produces a state of stress. For sufficiently shallow shells,  bending is not uniform, as the curvature variation localizes near the clamped side, see for instance Fig.~\ref{fig:experiment} (top) and the experimental and numerical results in \cite{brunetti2018}. Specifically, a small region is formed where the variation of Gaussian curvature --say $K_g$-- is localized. In Fig.~\ref{fig:experiment} such a region corresponds to a neighbourhood of the point $A$; indeed, the normals to the shell surface in the three points $A$, $B$ and $C$ identify a positive solid angle in the three-dimensional  unit sphere, distintive mark of a positive Gaussian curvature.
As the variation of Gaussian curvature implies the presence of membrane deformations, due to the Gauss Theorema Egregium, these regions are particularly interesting and were object of several studies, see for instance \citep{cerdaprl,contimaggi}.

%\begin{figure}
%	\centering
%	\includegraphics[width=0.5\linewidth]{figures/sideviewplus2.jpg}\\
%	\includegraphics[width=0.5\linewidth]{figures/dcone2.pdf}
%	\caption{Clamping a thin cylindrical shell (experiment): configuration with localized curvature. The variation of Gaussian curvature localizes within the region near the clamp, up to point $B$. The red arrows sketch the normals to the shell surface in the points $A$, $C$ and $D$.
%	}
%	\label{fig:experiment}
%\end{figure}

The relevant geometric parameters are sketched in Fig.~\ref{fig:geometry}; we call $\Omega:=[0,\ell]\times[-\varepsilon/2,\varepsilon/2]$ the shell planform, being $\ell$ the effective length of the part of the cantilever shell which does not undergo clamping and $\varepsilon$ its width.
Moreover, $h$ denotes the shell thickness and $\kappa_0$ its curvature in the $y$ direction. The total deepness of the initial configuration, see Fig.~\ref{fig:geometry}, can be expressed in terms of curvature being $\lambda=\varepsilon^2 \kappa_0/8+O(\kappa_0^3)$.  
The four geometric parameters ($h$, $\varepsilon$, $\kappa_0$ and $\ell$) are required to satisfy 
\begin{equation}
0 < h\ll \varepsilon, \qquad |\kappa_0| \lesssim \dfrac{1}{\varepsilon},\qquad \varepsilon \ll \ell,
\label{assumptions}
\end{equation}
corresponding respectively to a \emph{thin}, \emph{shallow} shell whose planform resembles a \emph{rod-like} body. As in the shallow regime the curvature scales as $1/\varepsilon$, we introduce the dimensionless parameter $k_0=\varepsilon \kappa_0=O(1)$.

The parameter $k_0$  determines the initial stress-free curvature and, therefore, the level of stress after clamping; for $k_0=0$ our problem becomes trivial. With reference to the axes chosen in Fig.~\ref{fig:geometry}, we have
\begin{equation}
\Kb_0=\dfrac{k_0}{\varepsilon}\,\mathbf{a}_y\otimes \mathbf{a}_y,\quad\text{ or }\quad
[\Kb_0]=\left[\begin{array}{cc}
0 & 0 \\ 
0 & k_0/\varepsilon
\end{array} \right],
\end{equation}
where $\mathbf{a}_y$ is the unit vector in the $y$ direction and $\Kb_0\in \mathrm{Sym}$ is the $2\times 2$ symmetric tensor defining the initial curvature. As the stress-free shape is cylindrical, its Gaussian curvature vanishes $K_{g0}=\det\Kb_0=0$.

\section{Dimensional reduction from the inextensible Koiter model}\label{sec:inex}

Before introducing the deduction from the FvK model, we begin with a  dimensional reduction starting from the nonlinear  model of Koiter's inextensible shells (see, \textit{e.g.}, chapter 10 in \cite{PGC}, or \cite{Miara}). The purpose is twofold: yielding some useful insights in a simpler context and  obtaining a first one-dimensional theory that allows to determine analytical solutions, a circumstance that is not affordable for the extensible model; the generalization to the more general case will then result more terse. A shell is said to be inextensible when its membrane deformations vanish almost everywhere on $\Omega$. The physical justification for such model is found in the limit $h\to 0$; since the ratio between the membrane and bending stiffnesses scales as $O(h^{-2})$, the relative cost of membrane deformations becomes increasingly high and these deformations tend to localize over set with vanishing area, curves (creases) or points (d-cones). We refer to \cite{muller} for more details.

Within the inextensible hypothesis, the shell stable configurations are found solving the following constrained minimization problem of the bending energy:
\begin{equation}
\min_{\Bb \in \mathcal{I}} \mathcal{E}_b(\Bb),\quad \mathcal{E}_b(\Bb):=\int_{\Omega} \dfrac{\Do}{2}\big(\Bb-\Bb_0\big)\cdot\big(\Bb-\Bb_0\big)\dd\Omega,
\label{mininext}
\end{equation}
where $
\mathcal{I}=\left\lbrace \Bb:\Omega\to \mathrm{Sym}, \curl\Bb=\mathbf{0},\,\det \Bb = 0\right\rbrace$.
The tensor fields $\Bb$ represents the finite curvature, $\Bb_0$ the stress-free finite curvature,  and  the constitutive tensor $\Do$ yields the bending stiffness.
%%The condition $\Bb \in \mathcal{I}$ means that the independent components, $(B_{xx},B_{yy},B_{xy})$, of the $2\times 2$ symmetric tensor field $\Bb$ satisfy 
%\begin{equation}
%\tR{B_{xx,y}=B_{xy,x},\quad  B_{xy,y}=B_{yy,x}, \quad B_{xx}B_{yy}=B_{xy}^2,}
%\label{curldet}
%\end{equation}}
%almost everywhere on $\Omega$. 
As it is known, the condition $ \curl\Bb=\mathbf{0}$ ensures the existence of a surface associated to the field $\Bb$. A classical approach, in order to satisfy this condition, is to seek solutions in the form
\begin{equation}
\Bb =  \widehat\Bb(\omega) = \nabla\nabla \omega,
\label{nablanablaw}
\end{equation}
for some scalar field $\omega:\Omega\to I\!\!R$. We remark that, other than being a potential to generate a vanishing curl tensor field, the scalar field has no direct physical meaning. However, when the shell is shallow, as in our  problem, the field $\omega$ can be interpreted as the \textit{transversal displacement field} $w$, (see \cite{calladine_1983}), as the finite curvature  can be approximated by the infinitesimal curvature $\Kb$
\begin{equation}
\Bb\simeq\Kb:=\nabla\nabla w.
\end{equation}
Once substituted $\Bb$ with $\Kb$ in \eqref{mininext}, the condition $\Bb\in \mathcal{I}$ translates into $\det \nabla\nabla w = w_{,xx} w_{,yy}-w_{,xy}^2=0$. 
%
%
% the minimization problem becomes:
%\begin{equation}\label{minw}
%\min_{w_{,xx} w_{,yy}=w_{,xy}^2} \mathcal{E}_b(w)=\int_{\Omega} \dfrac{\Do}{2}\big(\nabla\nabla w-\Bb_0\big)\cdot\big(\nabla\nabla w-\Bb_0\big)\dd\Omega,
%\end{equation} 
Hence, we use a simple \tR{Galerkin} method to deduce a one-dimensional rod model, and then provide the following  Ansazt for $w$:
\begin{equation}\label{displ}
w(x,y) = v(x)+\varepsilon\, k(x)\,\delta(y),
\end{equation}
where the function $\delta(y)$, expressing the $y$-dependence of the relevant fields in the problem, is given by
\begin{equation}\label{delta}
\delta(y)=\frac 12\left(\frac{y}{\varepsilon}\right)^2-\frac{1}{24}.
\end{equation}

\begin{remark} From \eqref{delta} we have  \begin{equation}
	\langle\delta\rangle=0, \quad \langle\delta'\rangle=0, \quad \langle\delta''\rangle=\varepsilon^{-2},
	\end{equation}
	where $\langle\psi\rangle:=(1/\varepsilon)\int_{-\varepsilon/2}^{\varepsilon/2} \psi(y)\,{\rm d}y$ represents the $y$-average value of the function $\psi(y)$. Using \eqref{displ}, one gets
	\begin{equation}\label{meaning}
	v(x) \equiv  \langle w(x,\cdot) \rangle, \;  k(x)\equiv\varepsilon \langle \partial_{yy} w(x,\cdot) \rangle.
	\end{equation}
	Hence, for any cross-section $x=\bar{x}$ of the shell, we can interpret $v(\bar{x})$ as the displacement in the $z$ direction of its center of mass (point $o$ in Fig.\ref{fig:geometry}) and $k(\bar{x})$ as the average of the dimensionless curvature  in the $y$-direction.
	
\end{remark}

\begin{remark} 
	Clearly more complex Ans\"{a}zte can be used. An easy improvement could be to increase the polynomial order of $\delta(y)$ to satisfy the boundary conditions for the bending moment, $\Do(\nabla\nabla(v+\varepsilon k \delta)-\Kb_0)$, along the sides $y=\pm \varepsilon/2$. However, we are only interested in the simplest possible choice allowing for the description of the Gaussian curvature along the rod axis $x$. The non-rigid micro-structure introduced using  \eqref{displ} is sufficient to our purposes. 
\end{remark}

\begin{remark}
	The functions $v(x)$ and $k(x)$ inherit, through \eqref{displ}, the regularity of $w$ and its boundary conditions. 
	Since $w$ is the shell transverse displacement,  when clamping the side $x=0$ we have
	$$
	w(0,y)=0,\; \partial_x w(0,y)=0, \quad \forall y\in[-\varepsilon/2,\varepsilon/2].
	$$
	Using \eqref{displ} we deduce $v(0)=0$, $v'(0)=0$, $k(0)=0$ and $k'(0)=0$. As the second derivatives of $w$ must be square integrable, then both $v$ and $k$ must belong to
	$$
	\mathcal{H} = \left\lbrace f\in H^2([0,\ell]), \; f(0)=0, \; f'(0)=0 \right\rbrace.
	$$ 
\end{remark}

\begin{remark}
	The minimization problem \eqref{mininext} does not involve external forces, because in the problem at hand there are not. Considering them is not straightforward, but manageable. The proposed approach is in the spirit of the so-called intrinsic elasticity (see, \emph{e.g.}, \cite{ciarletMardare}), for which the primary unknowns are the strain measures, instead of the displacement field, as in the classical approach. Adding forces within this framework requires to express the load potential in terms of strain --- in our case in terms of $\Kb$; to this purpose it is sufficient to have recourse to the Ces\`{a}ro-Volterra formula (see, \emph{e.g.}, \cite{cesaro}). Once the displacement is expressed as a linear functional of the curvature field, the resulting minimization problem will contain an additional linear term in $\Kb$.
\end{remark}

\noindent Using \eqref{nablanablaw} and \eqref{displ} we obtain the matrix field representing the approximated curvature, namely $\widetilde\Kb(v,k)=\nabla\nabla(v+\varepsilon k \delta)$. In matrix form, it reads
\begin{equation}
\widetilde\Kb(v,k)=\left(\begin{array}{cc}
v''(x)+\varepsilon k''(x)\delta(y) & \varepsilon k'(x)\delta'(y) \\ 
\cdot & \varepsilon k(x)\delta''(y)
\end{array} \right),
\label{Kreduced}
\end{equation}
where a prime indicates the derivative of a function with respect to its argument. 
When the functions $v$ and $k$ are varied in $\mathcal{H}$, $\widetilde{\Kb}(v,k)$ spans a subspace of 
$ L^2(\Omega,\mathrm{Sym})$ and, therefore, the reduced energy, namely 
$\widetilde{\mathcal{E}}_b(v,k):=\mathcal{E}_b(\widetilde\Kb(v,k))$, is finite. In particular, for isotropic materials, the reduced energy reads 
\begin{equation}\label{reducedEb}
\begin{aligned}
\widetilde{\mathcal{E}}_b(v,k)  =  \frac{D \varepsilon}{2}\int_{0}^{\ell}\left[ \bigl(v^{\prime\prime}\bigr)^2 + \frac{(k - k_0)^2}{\varepsilon^2} +\frac{2\nu(k - k_0) v^{\prime\prime} }{\varepsilon}   +\frac{(1-\nu)(k^\prime)^2}{6}  + \varepsilon^2 \frac{\bigl(k^{\prime\prime}\bigr)^2}{720}\right]\,\dd x,
\end{aligned}
\end{equation}
where $\Do\Kb=D[(1-\nu)\Kb+\nu\, (\mathrm{tr}\Kb)\,\Ib]$, with $D=Eh^3/12(1-\nu^2)$ the bending \tR{stiffness} in the $x$-direction, being \tR{$E>0$} the Young modulus, \tR{$-1\leq\nu\leq+1$ the 2D Poisson ratio} and $\Ib$ the identity tensor. 

As for the constraint in $\mathcal{I}$ requiring a curvature with a vanishing determinant, the Gaussian curvature of \eqref{Kreduced} turns out to be
\begin{equation}\label{Gauss}
\begin{aligned}
\det\widetilde\Kb(v,k)=\varepsilon\Big( v''(x)k(x)\delta''(y) \Big)+\varepsilon^2\Big( k(x)k''(x)\delta(y)\delta''(y)-k'(x)^2\delta'(y)^2 \Big).
\end{aligned}
\end{equation}
For $\widetilde{\Kb}$ to belong to $\mathcal{I}$ almost everywhere in $\Omega$, the field $k$ must vanish: a trivial solution due to the fact that $\widehat{\Kb}$ does not have sufficient degrees of freedom\footnote{More elaborate Ans\"atze with a finite number of terms would not help either.}. However, aiming at approximate solutions in the limit $\varepsilon/\ell\to 0$, we require only its $y$-average
\begin{equation}\label{meandetK}
\langle\det\widetilde\Kb(v,k)\rangle=\frac{1}{\varepsilon}k(x)v''(x)-\frac{1}{12}\,k'(x)^2,
\end{equation}
to vanish almost everywhere in $[0,\ell]$. 

Finally, we formulate our first rod model as the following constrained minimization problem
\begin{equation}\label{min1dinext}
\min_{(v,k)\in \mathcal{J}} \widetilde{\mathcal{E}}_b(v,k),
\end{equation}
where $\mathcal{J}=\left\lbrace (v,k)\in \mathcal{H}\times \mathcal{H},\; k\,v''=\varepsilon (k')^2/12 \right\rbrace$.
Being derived by \eqref{mininext}, this model will be referred in the following as the \emph{inextensible rod} model.

\section{Dimensional reduction from the F\"{o}ppl--von K\'{a}rm\'{a}n shell model}\label{sec:ex}

We discuss the derivation of a rod model starting from the assumption of a thin shallow shell satisfying the F\"{o}ppl--von K\'{a}rm\'{a}n equations. With respect to the previous section, the shell can undergo membrane deformations despite the fact that their cost (=stiffness) scales as $h$, whilst the cost of bending deformations, scaling as $h^3$, is sensibly smaller for thin shells.

If an initially curved shape of is considered, the model is often referred as  Marguerre--von K\'{a}rm\'{a}n's \cite{ciarletgMvK}.
For isotropic materials, it consists in finding the pair $(\varphi,w)$ such that
\begin{align}
& D\Delta\Delta (w-w_0)=[\varphi,w], \label{vKeq}\\
&(Eh)^{-1}\Delta\Delta\varphi=-\frac 12\big([w,w]-[w_0,w_0]\big),\label{vKcomp}
\end{align}
where, for two given scalar fields $a$ and $b$, $[a,b]$ denotes the Monge-Amp\`{e}re crochet\footnote{Specifically, in Cartesian coordinates we have $[a,b]=a,_{xx}b,_{yy}+a,_{yy}b,_{xx}-2a,_{xy}b,_{xy}$.}, and $\Delta$ is the Laplacian operator.

The fields $w$ and $w_0$ represent the displacements in the $z$-direction of the shell points in the current and stress-free configurations with respect to the flat reference configuration. As we have already discussed, in the limit of shallow shells, their second gradient gives the curvatures fields $\Kb=\nabla\nabla w$ and $\Kb_0=\nabla\nabla w_0$ for $w_0(x,y)=\varepsilon k_0 \delta(y)$.

The field $\varphi$ is the \textit{Airy stress function}. With respect to the case of inextensible shells, $\varphi$ is the additional field allowing to account for membrane stress and membrane deformations, respectively
\begin{equation}\label{airys}
\Nb(\varphi)=(\Delta \varphi)\Ib -\nabla\nabla\varphi, \quad \Eb(\varphi)=\Ao^{-1}\Nb(\varphi),
\end{equation}
and to define the membrane energy 
\begin{equation}
\mathcal{E}_m(\varphi):=\frac 12\int_\Omega\Nb(\varphi)\cdot\Ao^{-1}\Nb(\varphi),
\end{equation}
where  $\Ao=12\,\Do/h^2$ is the membrane stiffness. For more details on the derivation and meaning of the FvK equations we refer the interested reader to \citep{Audoly_2010}. In addition, we notice that general boundary conditions for the problem  \eqref{vKeq}--\eqref{vKcomp} are treated in several papers (see \cite{bfpv,ciarletgMvK}); in the following, we will consider only the special case corresponding to our problem.
%\footnote{$N_{xx}=\varphi,_{yy}$, $N_{yy}=\varphi,_{xx}$ and $N_{xy}=-\varphi,_{xy}$.}

In \cite{bfpv,bfv} we have shown that  eqs.  \eqref{vKeq}-\eqref{vKcomp} can be deduced  by enforcing the following mixed variational problem:
\begin{equation}
\min_{w\in \mathcal{W}}\max_{\varphi\in \mathcal{S}}\Fes(\varphi,w),
\end{equation}
where $\mathcal{W}$ and $\mathcal{S}$ are two suitable subsets of $H^{2}(\Omega)$ and the functional $\Fes(\varphi,w)$ is given by the splitting:
\begin{equation}\label{Fes}
\begin{aligned}
\Fes(\varphi,w)=\mathcal{E}_b(\nabla\nabla w)-\mathcal{E}_m(\varphi) +\frac 12\int_\Omega \Nb(\varphi)\cdot\big(\nabla w\otimes\nabla w-\nabla w_0\otimes\nabla w_0\big).
\end{aligned}
\end{equation}
In particular, for the case under consideration the Airy function $\varphi$ is required to vanish on $\partial\Omega$, so that $\mathcal{S}=H_0^2(\Omega)=\left\lbrace f\in H^2(\Omega), \; f=0, \; f,_n=0  \;\textrm{on}\;\partial\Omega\right\rbrace$.

Let us remark that both the bending and the membrane energy are quadratic and convex with respect to $w$ and $\varphi$. Last addend in \eqref{Fes} is the only term introducing \textit{non-linearities}; it does not have have any \textit{constitutive} character but couples the fields $w$ and $\varphi$, \emph{i.e.}, the bending and membrane problems.  

Dimensional reduction is achieved, once again,  via the \tR{Galerkin} method. In other words, we seek solutions for $w$ and $\varphi$ in the form
\begin{equation}\label{displphi}
w(x,y)=v(x)+\varepsilon k(x) \delta(y), \quad  \varphi(x,y)=f(x)\psi(y),
\end{equation}
with $\delta$ and $\psi$ given functions of the $y$-coordinate, see \eqref{delta} and \eqref{psidiy} respectively. The \emph{extensible rod} model is equivalent to the solution of the reduced min-max problem
\begin{equation}\label{minmax1d}
\min_{v\in \mathcal{H},k\in \mathcal{H}}\max_{f\in H^2_0(0,\ell)}\widetilde{\Fes}(f, v, k),
\end{equation}
with $\widetilde{\Fes}(f, v, k):=\Fes(f \psi, v+\varepsilon k \delta)$ the reduced action functional. The condition $f\in H_0^2(0,\ell)$ follows easily once noted that $\psi$ vanishes on the lateral boundaries $y=\pm\varepsilon/2$ (see \eqref{condbeta} below).

Specifically, we have used for $w$ the same Ansatz of the inextensible case, namely \eqref{displ} with $\delta$ as in \eqref{delta}. For the function $\psi$ expressing the $y$-dependance of the membrane fields, we choose the lowest-order polynomial satisfying
\begin{equation}\label{condbeta}
\psi(\pm\varepsilon/2)=0, \quad \psi'(\pm\varepsilon/2)=0, \quad \langle\psi\rangle=1,
\end{equation}
\emph{i.e.}
\begin{equation}\label{psidiy}
\psi(y)=30\left(\dfrac{y}{\varepsilon}\right)^4-15\left(\dfrac{y}{\varepsilon}\right)^2+\dfrac{15}{8}.
\end{equation}
Using \eqref{airys}$_1$ and \eqref{displphi}$_2$, this choice suffices to describe all the components of the membrane stress tensor:
\begin{equation}\label{Nij}
\begin{aligned}
&N_{xx}=\partial_{yy}\varphi=f(x)\psi''(y),\quad N_{yy}=\partial_{xx}\varphi=f''(x)\psi(y),\quad N_{xy}=-\partial_{xy}\varphi=-f'(x)\psi'(y). 
\end{aligned}
\end{equation}
Moreover conditions \eqref{condbeta} allow to satisfy the  boundary conditions $N_{yy}(x,y=\pm\varepsilon/2)=0$ and $N_{xy}(x,y=\pm\varepsilon/2)=0$  along the lateral sides of the shell.

For the reduced functional, in the case of isotropic materials, we obtain
%\begin{equation}
%\begin{aligned}
%&\widetilde{\Fes}(f,v,k)=\widetilde{\mathcal{E}}_b(v,k)-\widetilde{\mathcal{E}}_m(f)
%+\dfrac{v'_\ell(k_\ellf_\ell)-v'_0(k_0f_0)}{\varepsilon}\quad\medskip\\
% &\quad+\int_{0}^{\ell}\left(
%\frac{k'(kf)'}{84}+\frac{(k^2-k_0^2)f''}{56}-\frac{v''(kf)}{\varepsilon}\right)\,\dd x,
%\end{aligned}
%\end{equation}
\begin{equation}\label{isotropicF}
\begin{aligned}
&\widetilde{\Fes}(f,v,k)=\widetilde{\mathcal{E}}_b(v,k)-\widetilde{\mathcal{E}}_m(f)
+\int_{0}^{\ell}\left(
\frac{\varepsilon k'(kf)'}{84}+\frac{\varepsilon (k^2-k_0^2)f''}{56}+v'(kf)'\right)\,\dd x,
\end{aligned}
\end{equation}
where $\widetilde{\mathcal{E}}_b(v,k)$ is the reduced bending energy already computed in \eqref{reducedEb} and 
\begin{equation}\label{reducedEm}
\begin{aligned}
&\widetilde{\mathcal{E}}_m(f)= \frac{\varepsilon}{2Eh}\int_{0}^{\ell} \left( \frac{720}{\varepsilon^4}  \, f^2 +\frac{10}{7} \bigl(f^{\prime\prime}\bigr)^2  +\frac{240}{7\varepsilon^2}  [ (1 + \nu) \bigl(f^\prime \bigr)^2 + \nu f f^{\prime\prime} ] \right)\,\dd x
\end{aligned}
\end{equation}
is the reduced membrane energy.

\begin{remark}
	Since $ \langle\psi''\rangle=\psi(\varepsilon/2)-\psi(-\varepsilon/2)=0$, the mean value of $N_{xx}$ on the cross-section, \emph{i.e.} the axial stress in the resulting rod, vanishes. This is not the case in the model developed in \cite{guinot}. However, here the membrane stress $N_{xx}=f\psi''$ is  able to describe the zero-average stress distribution on the rod cross-section corresponding to a bending moment. %This will be shown in more detail in remark \ref{rembendingstiff}.
\end{remark}

\begin{remark}
	The Euler-Lagrange equation of $\widetilde{\mathcal{F}}(f,v,k)$ with respect to $f$ gives 
	\begin{equation}\label{comp1d}
	A[f]:=\frac{\varepsilon^4\,f''''}{504} -\frac{\varepsilon^2\,f''}{21}+f=-\frac{Eh\,\varepsilon^4}{720}\,\widetilde{K}_g(v'',k),
	\end{equation}
	where $\widetilde{K}_g(v'',k)=k \,v''/\varepsilon-(k')^2/28-k\, k''/42$ is another one-dimensional approximation of the shell Gaussian curvature. 
	Eq.~\eqref{comp1d} keeps track of the two-dimensional compatibility condition \eqref{vKcomp}. Given $(v'',k)$, Eq.~\eqref{comp1d} in terms of $f$ is analog 
	to the well known equation of a beam on an elastic ground \emph{\`{a} la}  Winkler (see, \emph{e.g.}, \cite{Timo}); formally its solution can be expressed as the convolution integral
	\begin{equation}\label{green}
	f^*(x)=-\frac{Eh\,\varepsilon^4}{720}\,\int_{0}^{x} G(x,s)\widetilde{K}_g(v'',k)(s)\,{\rm d}s,
	\end{equation}
	with $G(x,s)\in H^2_0(0,\ell)$ is the Green function of the fourth-order differential operator $A[f]$.
	
	Hence, the two-dimensional compatibility equation translates into a non-locality of the resulting rod model being formally equivalent to: 
	\begin{equation}\label{minmax1d2}
	\min_{v\in \mathcal{H},k\in \mathcal{H}}\max_{f\in H^2_0(0,\ell)}\widetilde{\Fes}(f, v, k)=\min_{v\in \mathcal{H},k\in \mathcal{H}}\widetilde{\Fes}(f^*, v, k).
	\end{equation}
	%	\texttt{\tR{Penso che questa sia sbagliata. Dovrebbe essere}}
	In particular, a root-finding calculation shows that the particular solutions of the differential operator $A[f]$ decay as $exp(-4.15\, x/\varepsilon)$. This implies that a concentrated variation of the Gaussian curvature of the order $O(1)$  decays to $O(10^{-2})$ at distance $\varepsilon$. Thus, the effective radius of non-locality is of the order $\varepsilon$.
\end{remark}

%\begin{remark}\label{rembendingstiff}
%CONTO STUPEFACENTE sulla bending stiffness 
%To this aim we suppose $v(x)=\bar{\chi}\, x^2 /2$, $k(x)=k_0$ and $f(x)=\bar{f}$. With this choice tha axial curvature $v''(x)=\bar{\chi}$ is constant and $v_\ell'-v_0'=\bar{\chi} \ell$. The action functional reduces to 
%\begin{equation}
%\widetilde{\Fes}(\bar{f},v,k_0)= \frac{D\ell}{2}\bar{\chi}^2-\frac{\ell}{2Eh}\frac{720}{\varepsilon^4}  \, \bar{f}^2
%\end{equation}
%with $\bar{f}=-E\, h\, \varepsilon^3 k_0 \,\bar{\chi}/720$ deduced from \eqref{comp1d}.
%Hence the energy is
%\begin{equation}
%\widetilde{\Fes}(\bar{f},v,k_0)= \left(D- \dfrac{E h}{720}\, \varepsilon^2 k_0^2\right) \frac{\ell}{2}\,\bar{\chi}^2
%\end{equation}
%The bending stiffness of the cross section shown in Fig.~\ref{fig:geometry} using the standard...
%\begin{equation}
%E I = E h\int_{-\varepsilon/2}^{\varepsilon/2} z^2(y) \,dy= \eta=\dfrac{E h}{720} \,\varepsilon^3 k_0^2
%\end{equation}
%\end{remark}

\section{Results}\label{results}

In this section, we solve the problem presented in Sec. \ref{problemsetup}, by adopting both the inextensible and  the extensible rod models; analytical results are possible just in the former case and this is why we  examine it first. In particular, our interest is to describe the region of localized curvature shown in Fig.~\ref{fig:experiment} estimating its width $d^*$, and the extremal values of curvature therein $\chi^*:=\max_{x\in[0,\ell]} v''(x)$ and $\chi^{**}:=\min_{x\in[0,\ell]} v''(x)$. For \tR{sake} of simplicity, in this section, we limit the analysis to the isotropic case.

\subsection{Analytical results: inextensible case}\label{inextloc}

We must solve the minimization problem \eqref{min1dinext}
with the reduced energy $\widetilde{\mathcal{E}}_b(v,k)$ given in \eqref{reducedEb}. Letting $\chi:=v''$ be the field of axial curvature and neglecting the term involving $(k'')^2$ in \eqref{reducedEb} (see remark \ref{remksec}), we face the following problem:
\begin{equation}\label{min1dinext2}
\begin{aligned}
\min_{\chi,k} \,\int_{0}^{\ell}\left( \chi^2 + \frac{(k - k_0)^2}{\varepsilon^2}+ \frac{2\nu(k - k_0) \chi }{\varepsilon}+\frac{(1-\nu)(k^\prime)^2}{6}  \right)
\end{aligned}
\end{equation}
with $\chi$ and $k$ constrained to satisfy $\chi k=\varepsilon (k')^2/12$.

\vspace{.3cm}
\begin{remark}\label{remksec}
	The coefficients weighting $(k')^2$ and $(k'')^2$ in \eqref{reducedEb} are respectively given by:
	\begin{equation}
	c_1= 2(1-\nu)\varepsilon \int_{-\varepsilon/2}^{+\varepsilon/2}\delta'(y)^2, \quad c_2=\varepsilon\int_{-\varepsilon/2}^{+\varepsilon/2}\delta(y)^2.
	\end{equation}
	Since $\langle\delta\rangle=0$, the Poincar\'{e}-Wirtinger inequality holds true and:
	\begin{equation}
	\int_{-\varepsilon/2}^{+\varepsilon/2}\delta'(y)^2\geq\frac{4\pi^2}{\varepsilon^2}\int_{-\varepsilon/2}^{+\varepsilon/2}\delta(y)^2.
	\end{equation} 
	Hence, we can estimate that $c_2/c_1\leq \varepsilon^2/(8(1-\nu)\pi^2)$, vanishing in the limit $\varepsilon/\ell\to 0$. 
\end{remark}
\vspace{.3cm}

%\begin{figure}	
%	\centering
%	\includegraphics[width=0.5\linewidth]{figures/cone}
%	\caption{Conical surface $\mathcal{C}$ representing the inextensibility constraint. The coordinates $c$ and $\vartheta$ allow to span the whole cone solving the constraint \eqref{comp1din}.  }
%	\label{conee}
%\end{figure}

\begin{figure}
	\begin{center}
		\small
		%	\hfill
		\subfloat[][]{
			\includegraphics[width=0.5\linewidth]{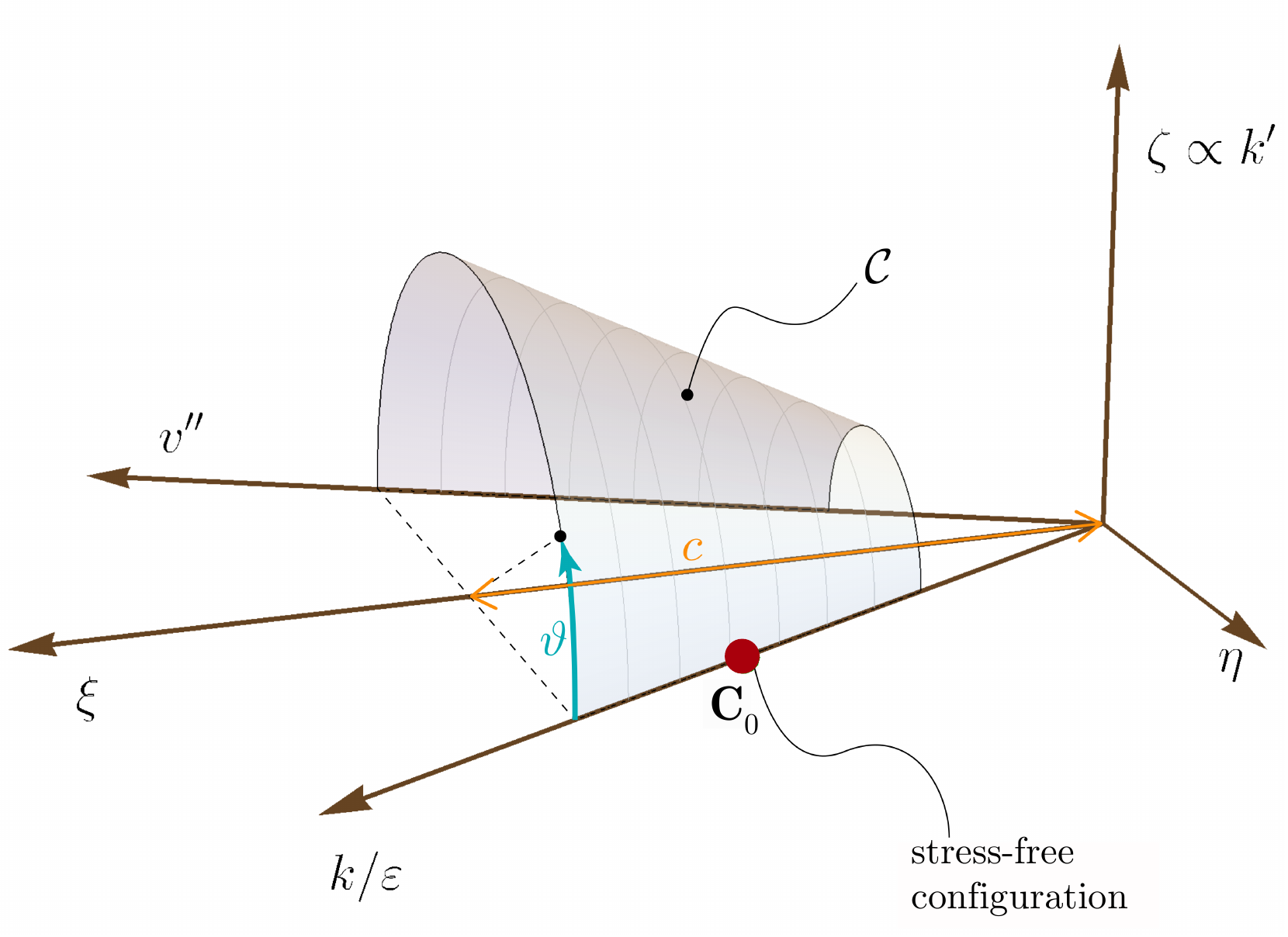}
		}
		%	\hfill
		\subfloat[][]{
			\includegraphics[width=0.45\linewidth]{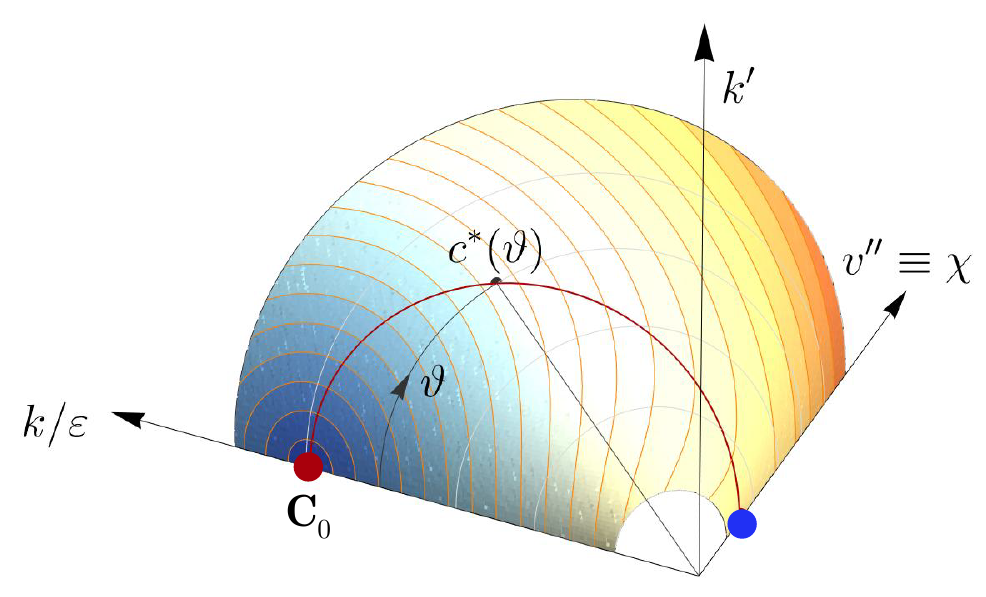}
		}
		\hfill
		\caption{ (a) Conical surface $\mathcal{C}$ representing the inextensibility constraint. The coordinates $c$ and $\vartheta$ allow to span the whole cone solving the constraint \eqref{comp1din}. (b) A curve over the cone (in red), passing through $\Cb_0$ and a point with optimal radial distance $c^*(\vartheta)$, see \eqref{cast}; this curve represents the set of configurations $(\chi,k,k')$ along the rod axis. The thin orange curves represent the locii of points having equal distance from $\mathbf{C}_0$ or, in physical terms, the configurations having the same density of bending energy. }
		\label{conee}
	\end{center}
\end{figure}

The problem \eqref{min1dinext2} is equivalent to
\begin{equation}\label{lesbc}
\min_{\mathbf{C}(x)\in\mathcal{C}} \int_{0}^{\ell}\Vert \mathbf{C}(x)-\mathbf{C}_0\Vert \, \dd x,
\end{equation}
with $\mathcal{C}$ the conic surface 
\begin{equation}\label{cone}
\mathcal{C}:=\left\{ (\xi, \eta, \zeta) \;\Big|\;  \xi^2=(1+\nu)\left(\eta^2+\zeta^2\right)/(1-\nu)  \right\},
\end{equation}
and $\mathbf{C}(x)$, $\mathbf{C}_0$  the points of coordinates:
\begin{equation}\label{changecoord}
\mathbf{C}(x)=\left( 
\begin{array}{l}
\xi=\sqrt{1+\nu}\left({k(x)}/{\varepsilon}+ \chi(x) \right)/2\\
\eta=\sqrt{1-\nu}\left({k(x)}/{\varepsilon}- \chi(x)\right)/2\\
\zeta=\sqrt{1-\nu}\, (k'(x))^2/(2\sqrt{3})
\end{array}\right), \qquad \mathbf{C}_0=\left( 
\begin{array}{l}
\sqrt{1+\nu}\left({k_0}/{\varepsilon} \right)/2\\
\sqrt{1-\nu}\left({k_0}/{\varepsilon}\right)/2\\
0
\end{array}\right),
\end{equation}
%and
%\begin{equation}
%\mathbf{C}_0=\left( 
%\begin{array}{l}
%\sqrt{1+\nu}\left({k_0}/{\varepsilon} \right)/2\\
%\sqrt{1-\nu}\left({k_0}/{\varepsilon}\right)/2\\
%0
%\end{array}\right),
%\end{equation}
this latter corresponding to the stress-free configuration. 
Under the proposed change of coordinates, see \cite{Hamouche2017}, minimizing the bending energy under the inextensibility constraint is reformulated as the search of a sequence of points $\mathbf{C}(x\in[0,\ell])$ lying  on the cone $\Cc$ having minimal (total) distance from the target point $\mathbf{C}_0$ (see Fig.~\ref{conee}).

It easily seen that the condition $\mathbf{C}(x)\in\mathcal{C}$ translates the inextensibility constraint or, in other words, the condition 
\begin{equation}\label{comp1din}
k(x)\,\chi(x)=\frac{1}{12}\varepsilon (k'(x))^2
\end{equation}
for the cross-section average of the Gaussian curvature to vanish.

However, having this geometric understanding of the problem, it is clear that for a smooth parametrization of the inextensibility constraint, we need an angular coordinate. Specifically, we
use the cone coordinates $(c,\vartheta)$ shown in  Fig.~\ref{conee}: any point of the cone can be written in the form
\begin{equation}\label{change}
\widetilde{\mathbf{C}}(c,\vartheta)= c \left( 1, \sqrt{\frac{1-\nu}{1+\nu}}\,\cos\vartheta, \sqrt{\frac{1-\nu}{1+\nu}}\,\sin\vartheta\right),
\end{equation}
for some choice of the angular anomaly $\vartheta$ and of the curvature\footnote{The coordinate $c$ is actually proportional to the mean curvature $(\chi+k/\varepsilon)/2$.} $c$. Of course, there is a correspondence between $\Cb(x)$ and $\widetilde{\Cb}(c,\vartheta)$, which has to been determined.
For a given point $x$, the compatibility constraint \eqref{comp1din} reveals how $\chi$, $k$ and $k'$ have to be related in order to be admissible at that point. If we adopt the parametrization of the cone in terms of $(c,\vartheta)$, we have to put in correspondence $\chi(x)$ and  $k(x)$  with some auxiliary functions $\widetilde{\chi}(c,\vartheta)$ and $\widetilde{k}(c,\vartheta)$.  In other words, we need to find a correspondence between the physical point $x$ and the coordinates $(c,\vartheta)$, in order to find the map $x\leftrightarrow\widetilde{x}(c,\vartheta)$. From a geometrical point of view, this means that the admissible configuration of a point of the rod may be visualized as a point over the cone $\widetilde{\mathbf{C}}(c,\vartheta)$;  
a sequence of points (a curve) laying on the cone then represents the set of the admissible configurations of the whole rod. A way to minimize the bending energy, is to require that each point of such a curve has minimum distance $d(c,\vartheta)=\Vert \widetilde{\mathbf{C}}(c,\vartheta)-\mathbf{C}_0\Vert$ from the point corresponding to the stress-free configuration. It is not difficult to see that
\begin{equation}\label{conedis}
d(c,\vartheta) = \frac{2c^2}{1+\nu}+\frac{k_0^2-2\,k_0 c\varepsilon\sqrt{1+\nu}}{2\varepsilon^2}-\frac{c(1-\nu)\cos\vartheta}{\sqrt{1+\nu}}\, \frac{k_0}{\varepsilon}.
\end{equation}
Moreover,  the axial and transverse curvatures can be expressed as
\begin{equation}\label{chitilde}
\widetilde{\chi}(c,\vartheta)=\frac{c(1-\cos\vartheta)}{\sqrt{1+\nu}}, \quad \widetilde{k}(c,\vartheta)=\frac{\varepsilon c(1+\cos\vartheta)}{\sqrt{1+\nu}},
\end{equation}
respectively.

Requiring the distance \eqref{conedis} to be minimal with respect to $c$ implies 
\begin{equation}\label{cast}
c=c^*(\vartheta)=\frac 14\sqrt{1+\nu}\Big( 1+\nu+(1-\nu)\cos\vartheta \Big)\,\frac{k_0}{\varepsilon}.
\end{equation}
Fig. \ref{conee} (b) is instrumental to understand the geometrical meaning of the procedure we intend to adopt. For a fixed anomaly $\vartheta$, we determine the optimal $c^*(\vartheta)$, \textit{i.e.}, the optimal value of the radial coordinate (or rather the mean curvature) in order to minimize the bending energy; then we need to determine the curve (red in Fig. \ref{conee} (b)) through $c^*(\vartheta)$, still lying over the cone, whose points locally minimize the distance from $\Cb_0$, representing the stress-free configuration. It is worth noticing that such a procedure requires the \textit{local} minimization of the distance $d(c,\vartheta)$ via \eqref{cast}. This is a condition sufficient, but not necessary, to minimize \eqref{minmax1d2}, \emph{i.e.}, the total distance to $\mathbf{C}_0$. Solutions minimizing the total distance without necessarily satisfying \eqref{cast} in the whole domain $[0,\ell]$ could be possible, in principle.

On substituting \eqref{cast} in \eqref{chitilde}, we get:
\begin{equation}\label{cap}
\begin{aligned}
& \widehat{\chi}(\vartheta)=\widetilde{\chi}\big( c^*(\vartheta),\vartheta\big)=\frac{k_0 (1-\cos\vartheta)}{4\varepsilon}\Big( 1+\nu+(1-\nu)\cos\vartheta \Big),\\
& \widehat{k}(\vartheta)=\widetilde{k}\big( c^*(\vartheta),\vartheta\big)=\frac{k_0 (1+\cos\vartheta)}{4}\Big( 1+\nu+(1-\nu)\cos\vartheta \Big).
\end{aligned}
\end{equation}
At this point, if  knew the map $\vartheta=\widehat{\vartheta}(x)$, we would be in position to obtain the solution of the problem, determining the unknowns $\chi(x)=\widehat{\chi}\big(\widehat{\vartheta}(x)\big)$ and $k(x)=\widehat{k}\big(\widehat{\vartheta}(x)\big)$. This is our next goal.

The compatibility equation suggests the one-to-one mapping $x\mapsto \vartheta$ between the spatial variable and the angular variable on the cone which is the key to solve  the problem. Indeed, from $(k')^2=12 k \chi/\varepsilon$ we obtain
\begin{equation}\label{xpth}
\frac{\dd \widehat x}{\dd \vartheta}=\frac{1}{k'}\frac{\dd \widehat{k}}{\dd \vartheta}=\frac{\dd \widehat k(\vartheta)}{\dd \vartheta}\frac{\sqrt{\varepsilon}}{\sqrt{12\,	\widehat k(\vartheta)\widehat \chi(\vartheta)}}
\end{equation}
%=\frac{-\big(1+(1-\nu)\big)\cos\vartheta}{\sqrt{3}\big( 1+\nu+(1-\nu)\cos\vartheta \big)}.
valid for $0\le\vartheta<\pi$.
Inserting \eqref{cap} and integrating, we deduce
\begin{equation}\label{xtheta}
\widehat{x}(\vartheta)=\varepsilon\,\frac{ 
	(2-\sqrt{\nu})\pi-2\vartheta+2\sqrt{\nu}\tan^{-1}\left(\sqrt{\nu}\tan\frac{\vartheta}{2}\right)}{2\sqrt{3}}
\end{equation}
It is easily checked that, for $0<\nu<1$ and $\vartheta\in [0,\pi)$, $\widehat{x}'$ is strictly negative and $x$ maps the set $(0,\pi)$ into $(d^*,0)$ monotonically with $d^*(\nu)=\pi\varepsilon\,(2-\sqrt{\nu})/(2\sqrt{3})$.

\begin{figure}
	\centering
	\includegraphics[width=0.5\linewidth]{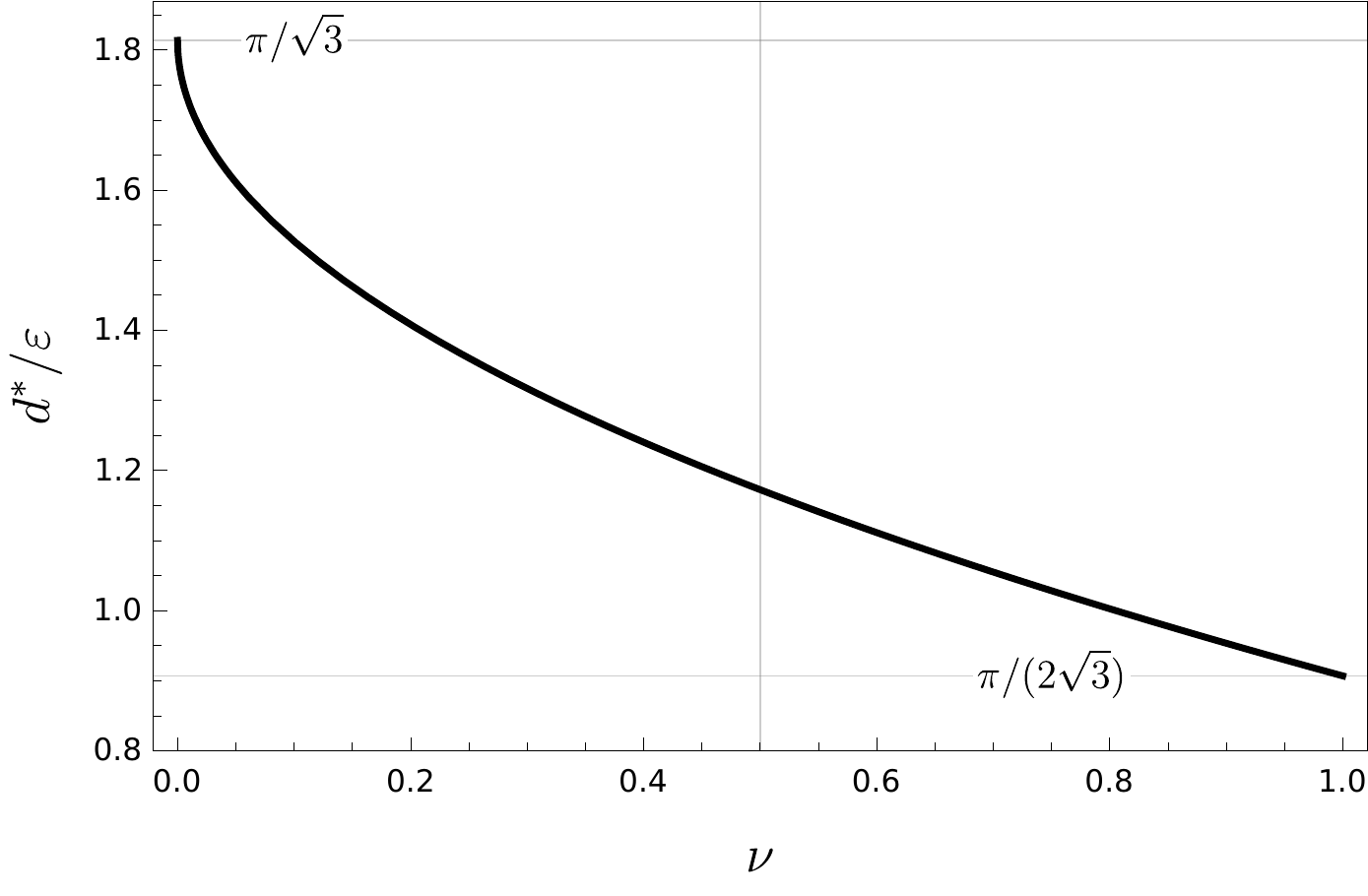}
	\caption{Size of the region where the curvature localizes: $d^*/\varepsilon$ as a function of the Poisson \tR{ratio}.}
	\label{fig:destimate}
\end{figure}

\begin{figure}
	\begin{center}
		\small
		%	\hfill
		\subfloat[][]{
			\includegraphics[width=0.5\linewidth]{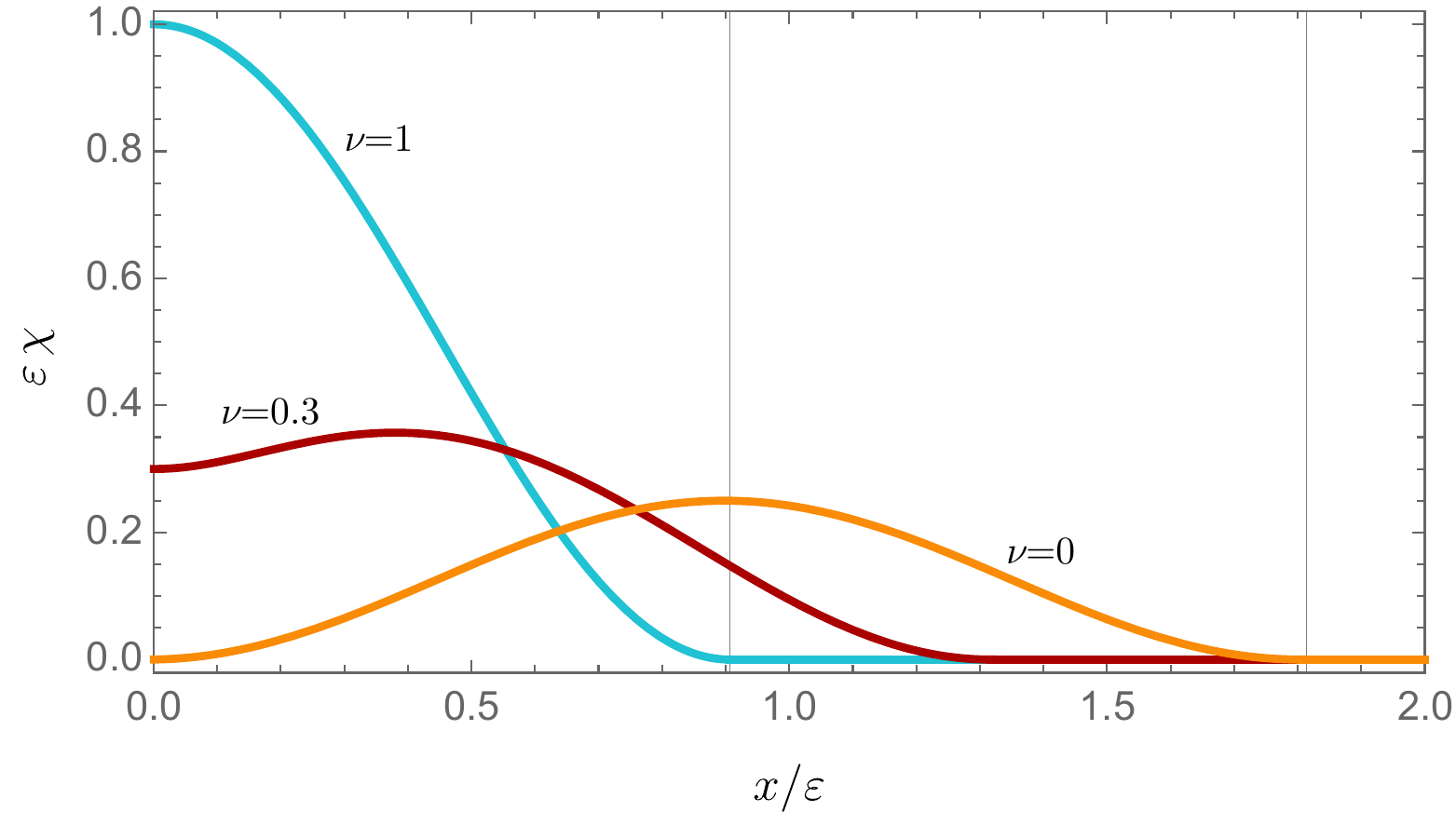}
		}
		%	\hfill
		\subfloat[][]{
			\includegraphics[width=0.5\linewidth]{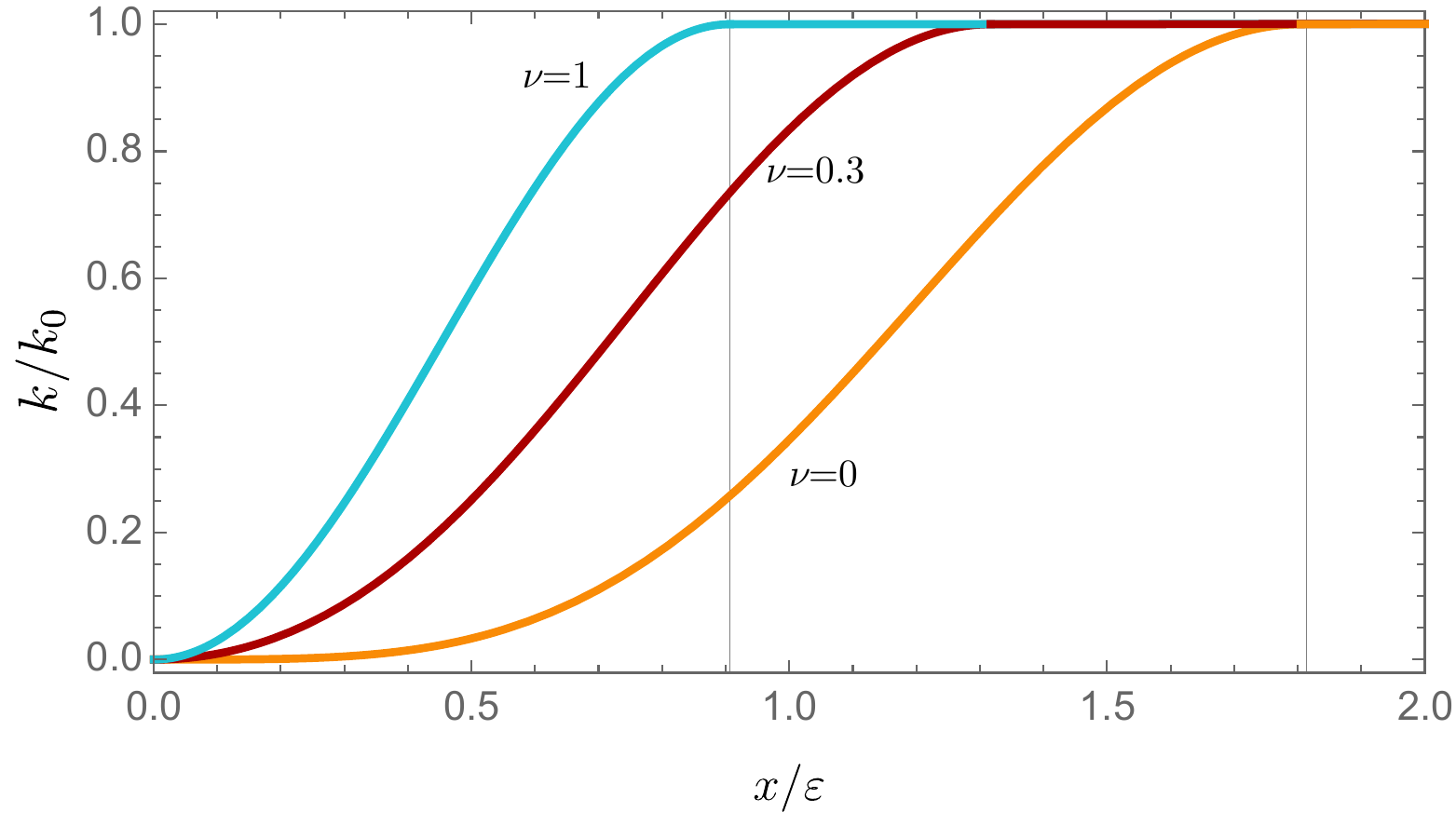}
		}
		\hfill
		\caption{Fields of axial curvature $\varepsilon\chi$ (a) and transversal curvature $k/k_0$ (b) for several values of the Poisson ratio.}
		\label{fig:chikinext}
	\end{center}
\end{figure}

The inverse function of \eqref{xtheta}, say $\widehat{\vartheta}(x)$,  allows to determine $\chi(x)$ and $k(x)$ from \eqref{cap}. This solution is valid until the point $\mathbf{C}_0$ in Fig. \ref{conee} is reached for $x=d^*$. Indeed, for $x>d^*$ the distance $\Vert \mathbf{C}(x)-\mathbf{C}_0\Vert$ is minimized by remainining in the same point $\mathbf{C}(x)\equiv\mathbf{C}(d^*)=\mathbf{C}_0$. 
Thus $d^*=O(\varepsilon)$ is actually the size of the region where the curvature localizes; the ratio $d^*/\varepsilon$ is plotted against $\nu$ in  Fig.~\ref{fig:destimate}. 
Unfortunately, the inverse function $\widehat{\vartheta}(x)$ of \eqref{xtheta} cannot be analytically determined for arbitrary values of $\nu$; however the problem of its numerical determination is well-posed since $\widehat{x}$ is strictly monotone. The solutions for the axial and transverse curvatures are plotted in Fig.~\ref{fig:chikinext} for some values of the Poisson ratio.

%
%\begin{equation}\label{dxg}
%\dd x=\frac{\sqrt{\varepsilon}\,\dd k}{\sqrt{12 \, k \,\ov \chi(k)}};
%\end{equation} 
%here we have removed the ambiguity of the sign because $k(x)$ is a function monotonically increasing from $0$ to $k_0$. 
% 

\begin{figure}
	\centering
	\includegraphics[width=0.5\linewidth]{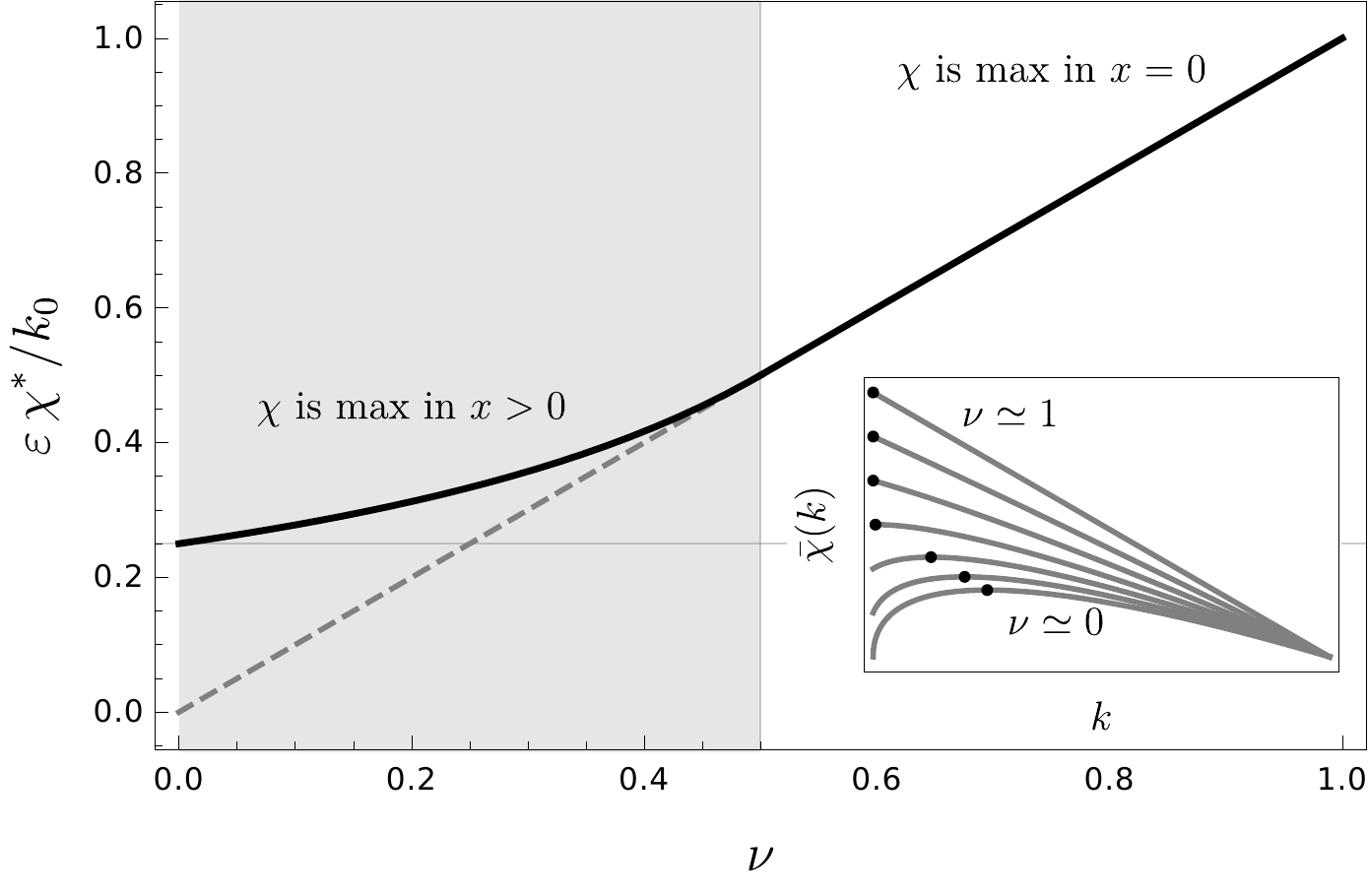}
	\caption{Maximum value $\chi^*$ of the axial curvature, normalized with respect to the initial transversal curvature $k_0/\varepsilon$, as a function of the Poisson \tR{ratio}. For $0\leq\nu<1$ (gray region), the maximum is attained in $x>0$, whilst it is in $x=0$ whenever $\nu>0$. In the inset the function $\ov{\chi}(k)$ is represented, showing the points $k^*$ where the maximum is attained.}
	\label{chimax}
\end{figure}

Although $\widehat{x}(\vartheta)$ cannot be analytically determined, in general, closed forms of  the maximal value of the axial curvature can be achieved. To this end, we eliminate $\vartheta$ from \eqref{cap} and introduce the map
\begin{equation}\label{vsg}
k\mapsto\ov \chi(k)=\frac{\nu k_0-2k+\sqrt{\big( 4k (1-\nu)+k_0\nu^2\big )k_0}}{2\varepsilon}.
\end{equation}
The maximum  is attained for $k^*$, solution of the stationarity condition $\partial_k\ov \chi(k)=0$, yielding
\begin{equation}
k^*= \max\left\{ 0,\frac{1-2\nu}{4(1-\nu)}\,k_0 \right \}.
\end{equation}
Here, we have used the fact that $k$ is a monotonically increasing positive function, whose codomain is $[0,k_0]$. Finally, the maximum value of the axial curvature scales as $O(\varepsilon^{-1})$   being
\begin{equation}\label{chias}
\chi^*=\ov \chi(k^*)=\dfrac{k_0}{\varepsilon} \times \left\{
\begin{array}{ll}
\dfrac{1}{4(1-\nu)} & \mbox{if $0< \nu< 1/2$}   \\[0.3cm]
\nu & \mbox{if $1/2\leq\nu\leq 1$}
\end{array}
\right.
\end{equation}
which is plotted in Fig.~\ref{chimax} against positive Poisson ratios. It is easily seen that the axial curvature is always non negative; and therefore, the minimum value of the curvature is $\chi^{**}=0$ for any $\nu$. 

\vspace{.3cm}
\begin{remark}
	The solution found implies a localization of the axial rod curvature that tends to a Dirac delta distribution in the limit $\varepsilon/\ell\to 0$. Indeed, $\chi^*=O(\varepsilon^{-1})$ over a region $d^*=O(\varepsilon)$, the integral
	\begin{equation}\label{vprime}
	v'(\ell)=\int_{0}^{\ell} \chi(x) \dd x=\int_{\pi}^{0} \widehat{\chi}(\vartheta) d\vartheta=\dfrac{(1+\nu)\pi k_0}{8\sqrt{3}},
	\end{equation} 
	being finite and independent of $\varepsilon$.
\end{remark}

\subsection{Numerical results for the extensible cases}

The solutions of the extensible rod are found as saddle points, see 
\eqref{minmax1d}, of the functional $\mathcal{F}(v,k,f)$ given in  \eqref{isotropicF}. As a numerical procedure is necessary to this end, we use a standard finite element method. The domain $[0,\ell]$ is discretized into $n$ elements with a mesh suitably refined near the clamp $x=0$. Since all the fields belongs to $H^2$, we choose, for all of them, Langrange polynomials of order 3 ensuring the inter-element continuity of their values and their first derivatives. In every node of the mesh we have $3\times2$ degrees of freedoms (total size of the problem $6 n+6$ scalar unknowns). Storing in the vector $\mathbf{q}$ the degrees of freedom relative to $v$ and $k$ and in the vector $\mathbf{f}$ the ones relative to $f$, the action functional \eqref{isotropicF} is written as
\begin{equation}\label{discreteF}
\mathcal{F} \simeq \frac{1}{2}\mathbb{K}\mathbf{q}\cdot\mathbf{q}-k_0 \mathbb{L}\mathbf{q}-\frac{1}{2}\mathbb{H}\mathbf{f}\cdot\mathbf{f}+\mathbb{C}\mathbf{q}\mathbf{q}\cdot\mathbf{f}-k_0^2 \mathbb{M}\mathbf{f},
\end{equation}
with $\mathbb{K}$ and $\mathbb{H}$ positive definite second order tensors, $\mathbb{L}$ and $\mathbb{M}$ vectors and $\mathbb{C}$ the third order tensor responsible for the coupling between membrane and bending problems. We compute once for all these tensors avoiding the reassembling of the stiffness matrices even if is a nonlinear problem. The saddle point satisfying \eqref{minmax1d} is found by iteratively finding the root of the following system:
\begin{equation}\label{deqns}
\left\lbrace
\begin{array}{l}
0=\partial_\mathbf{q}\mathcal{F} = \left(\mathbb{K}+2 \mathbb{C}^\top\mathbf{f}\right)\mathbf{q}-k_0 \mathbb{L},\medskip\\
0=\partial_\mathbf{f}\mathcal{F} = -\mathbb{H}\mathbf{f}+\mathbb{C}\mathbf{q}\mathbf{q}-k_0^2 \mathbb{M},
\end{array}
\right.
\end{equation}
Clearly, the system can have several solutions depending on the initial guess $(\mathbf{q}_i,\mathbf{f}_i)$, cfr. section \ref{sec:multistab}. To follow the equilibrium branch  relative to the curvature localization shown in Fig.~\ref{fig:experiment} suffices to start from  $\mathbf{q}_i=\mathbf{0}$, $\mathbf{f}_i=\mathbf{0}$.

As a benchmark solution for our reduced rod models, we numerically solve the FvK shell equations with the boundary conditions provided by the problem at hand. 
To this aim we resort to the code provided by the FEniCS shells project \cite{fenics,fenicsshells} which implements a standard di\-spla\-ce\-ment-based FE procedure (that is, the solution is found by minimizing the shell elastic energy).
The domain $[0, \ell] \times [-\varepsilon/2, \varepsilon/2]$ has been discretized with both structured and unstructured meshes suitably refined near the clamp $x=0$ so that the mesh size is smaller than the shell thickness, while membrane locking is avoided by appropriately choosing the discrete spaces for in-plane and transverse displacements.
More in detail, we used standard Lagrangian elements for both of them, weakly enforcing $C^1$-continuity of the piecewise continuous transverse displacement by penalizing the jump of the normal component of its gradient through the element facets.
We chose the penalty parameter to be of the order of the norm of the bending stiffness tensor.

\subsection{Comparisons}

\begin{figure}
	\begin{center}
		\small
		%	\hfill
		\subfloat[][]{
			\includegraphics[scale=0.51]{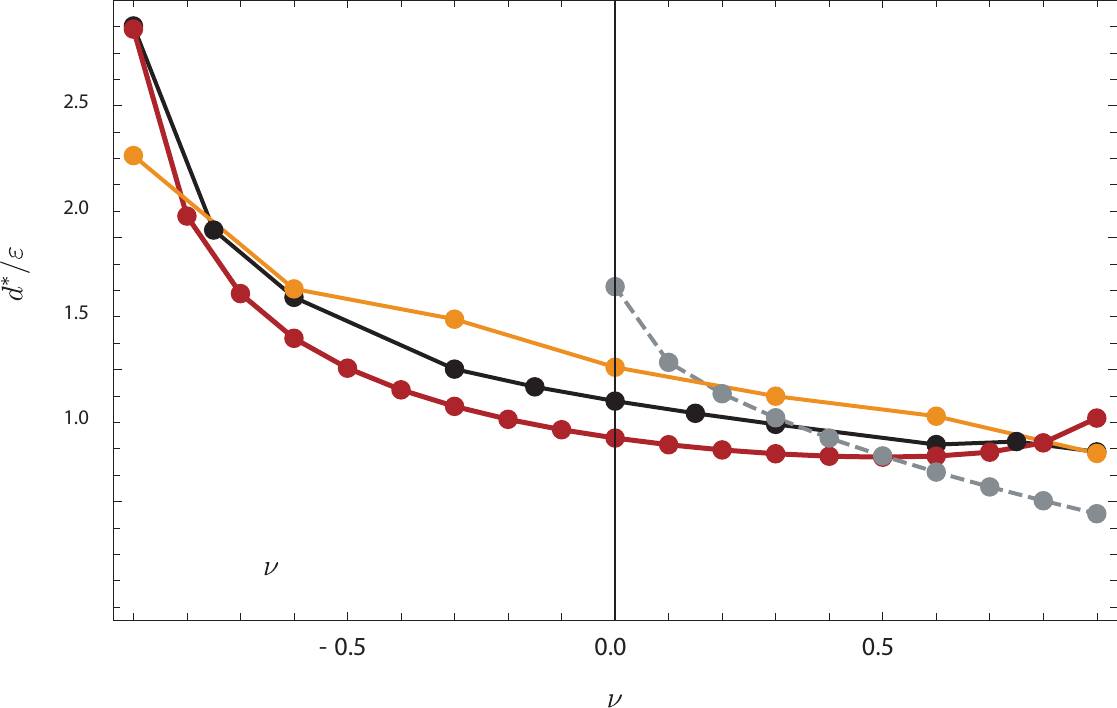}  \label{fig:compchidstara}		}
		%
		%	\hfill
		\subfloat[][]{		\includegraphics[scale=0.5]{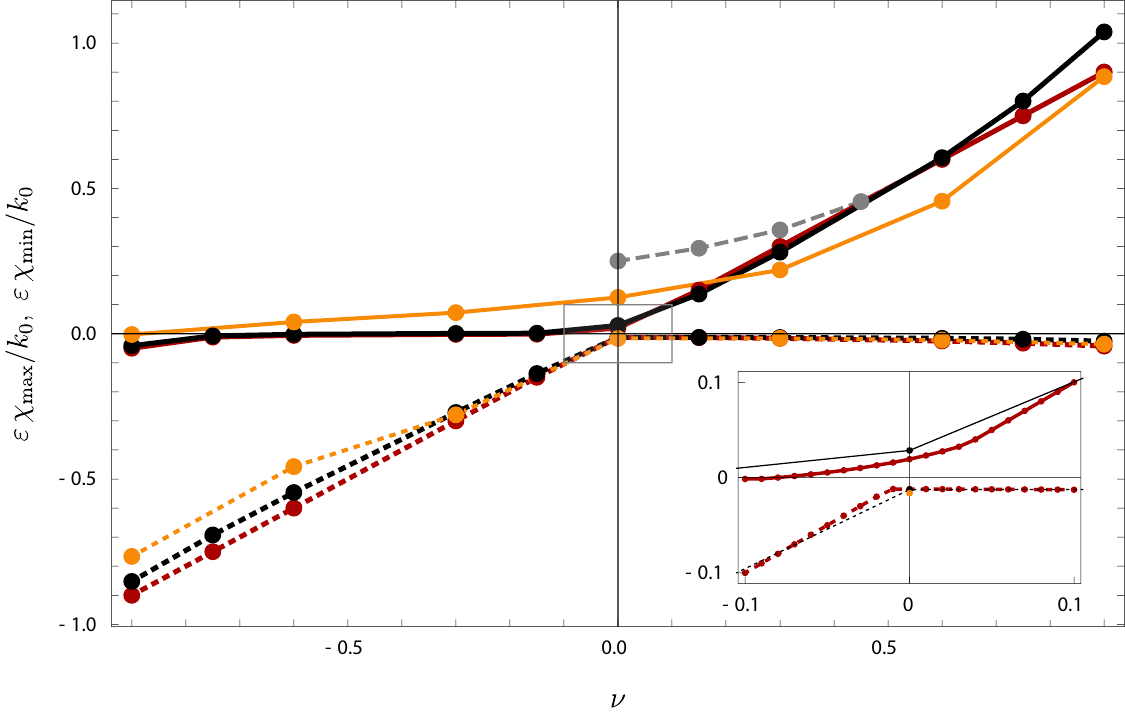}\label{fig:compchidstarb}	}
		\hfill
		\caption{Plot of the size $d^*/\varepsilon$ of the localization region (a) and of the maximal  and minimal axial curvatures (b) as functions of the Poisson \tR{ratio}: inextensible rod (gray), extensible rod (red, continuous line for the maximum $\chi^*$ and dashed line for the minimum $\chi^{**}$), FvK shell (black,  continuous line for the maximum $\chi^*$ and dashed line for the minimum $\chi^{**}$)\tR{, fully non-linear Naghdi shell model (orange, continuous line for the maximum $\chi^*$ and dashed line for the minimum $\chi^{**}$)}; the minimum value for the inextensible rod is zero.}
	\end{center}
	%\label{fig:compchidstar}
\end{figure}	

We compare the results obtained by the FvK shell model, assumed as a benchmark, with the inextensible and extensible rod models derived in sections \ref{sec:inex} and \ref{sec:ex}. In the numerical simulations we have chosen $k_0=1$, $\varepsilon=20\,h$ and $\ell=20\,\varepsilon$ this last choice being irrelevant as far as $\ell\gg \varepsilon$. Indeed, the localization of curvature happens within a distance $(2\div 3)\, \varepsilon$ from the clamp and this is actually the only region where we have plotted the relevant fields.
Results are independent of the Young modulus but do depend on the Poisson ratio.

Figs.~\ref{fig:compchidstara} and \ref{fig:compchidstarb} plot the maximal and minimal axial curvature and the size of localization region as estimated by the three models under consideration for the admissible range of Poisson ratio $\nu\in(-1,1)$. The inextensible rod model results are limited to the case $\nu\ge 0$: negative values of the Poisson ratio are in principle possible, but the geometric construction presented in \ref{inextloc} would require to consider the singular point corresponding to the vertex of the cone, an analytical obstacle that we exclude for sake of simplicity. We used a black color to label the benchmark FvK shell model, gray and dark-red curves to indicate the inextensible and extensible rod models, respectively. \tR{As a further benchmark, we performed a FE-based analysis with the fully non-linear Naghdi shell model \cite{fenicsshells} (orange curves); we conclude that all the relevant estimates are well captured by our theory.}
We see that the sup-norm of the curvature field is very well estimated by  the simple formula
\begin{equation}
\Vert \chi \Vert_{\infty} = \max(\vert\chi^*\vert,\vert\chi^{**}\vert)\simeq\frac{k_0}{\varepsilon}\,\vert\nu\vert.
\end{equation}
We recall, see \eqref{chias}, that this can be obtained as the point on the cone axis $k=0$, $k'=0$ having minimal distance form the stress-free configuration $\mathbf{C}_0$. The inset in Fig. \ref{fig:compchidstarb} reveals that the maximum and minimum values of the curvature, considered as functions of $\nu$, never intersect: thus, the axis is always bent,  even close to $\nu=0$.
In general, we remark a good agreement of both the rod models with the two-dimensional results. In all three cases, the maximum of axial curvature scales as $1/\varepsilon$ whilst the localization size scales as $\varepsilon$; hence, the rod models are able to catch the macroscopic deformation of the rod,  independently of the cross-section dimension $\varepsilon$. 
The inextensible model overestimates the maximal curvature for vanishing values of the Poisson ratio: in this case the interplay between axial and transverse curvature, already constrained by the inextensibility hypothesis, is further limited since the coupling terms $\Do_{12}$ and $\Do_{21}$ of the Voigt representation of $\Do$ vanish with $\nu$.

\begin{figure}
	\begin{center}
		\small
		%	\hfill
		\subfloat[][]{
			\includegraphics[scale=0.5]{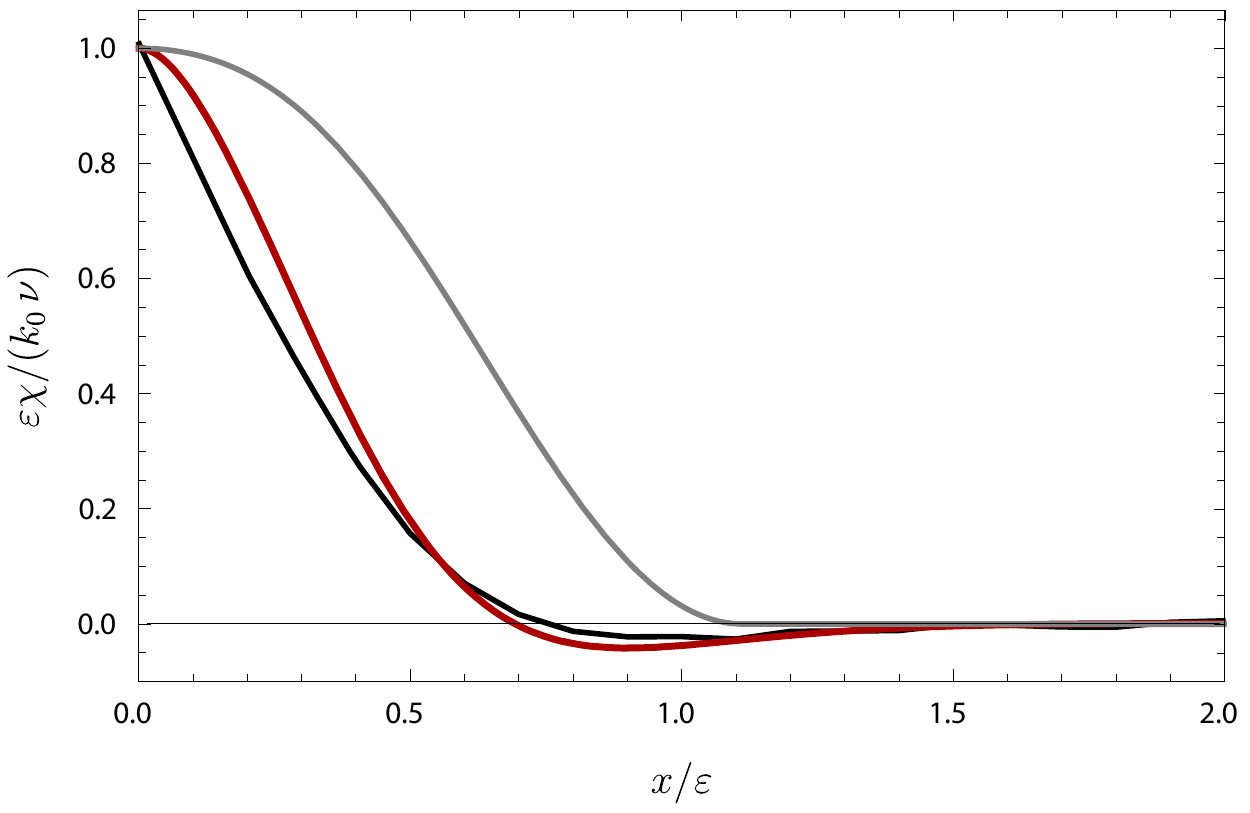}
			\label{curvature1Da}
		}
		%
		%	\hfill
		\subfloat[][]{
			\includegraphics[scale=0.5]{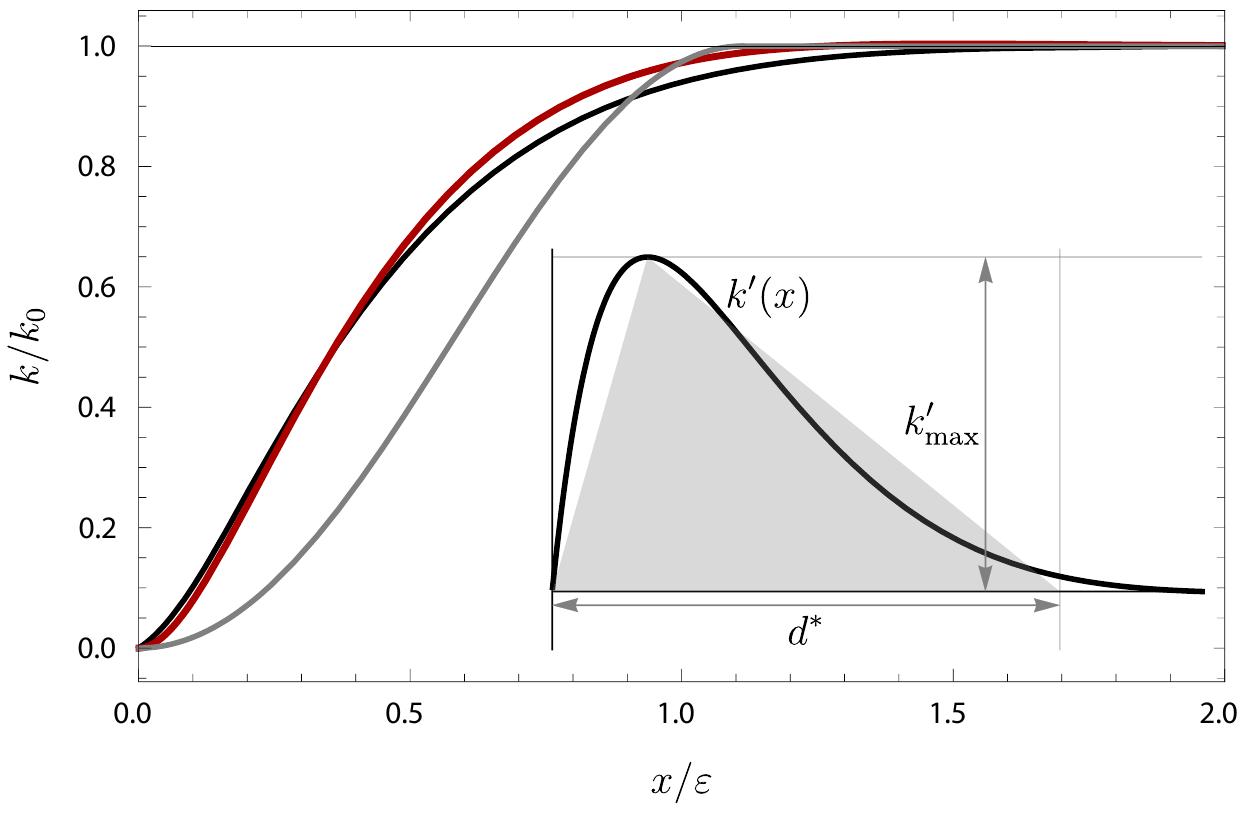}
			\label{curvature1Db}
		}
		\hfill
		\caption{Plot of $\varepsilon \chi/k_0\nu$ (a) and $k/k_0\nu$ (b), according to the one-dimensional theory (red), the FvK model (black), the inextensible model (gray).}
	\end{center}
\end{figure}	

For $\nu=0.6$, Figs. \ref{curvature1Da} and  \ref{curvature1Db}  plot the spatial distributions of the  axial  $\chi(x)=v''(x)$ and transverse $k(x)$ curvatures\footnote{For the FvK shell the axial and transverse curvatures are obtained from the displacement field $w(x,y)$ through \eqref{meaning}.} within the localization region. While the extinction length $d^*$ for the inextensible case has been obtained in closed form in Sect.~\ref{inextloc}, both the curvatures of the FvK shell and of the extensible rod exponentially decay towards their asymptotic values $\chi(x\to\ell)\simeq 0$ and $k(x\to\ell) \simeq k_0$. For both these models the size $d^*$ of the localization region has been estimated approximating the area subtended to the graph of $k'(x)$ by a triangle having height $k'_{\max}$, namely
\begin{equation}
k_0 \simeq \int_{0}^{\ell} k'(x)\,\dd x\simeq   \dfrac{d^* k'_{\max}}{2} \;\Rightarrow \; d^*\simeq \dfrac{2 k_0}{k'_{\max}},
\end{equation}
as shown in the inset of Fig.~\ref{curvature1Db}.

%Both the non-local model and inextensible model capture very well the maximum amplitude of the curvatures, the extinction length, having order $\varepsilon$, and then the localization. Discussion about the minimization of the bending energy without taking into account the compatibility is postponed to Sec. \ref{discuss}.

\begin{figure}
	\begin{center}
		%	\hfill
		%	\subfloat[][]{
		\includegraphics[width=0.5\linewidth]{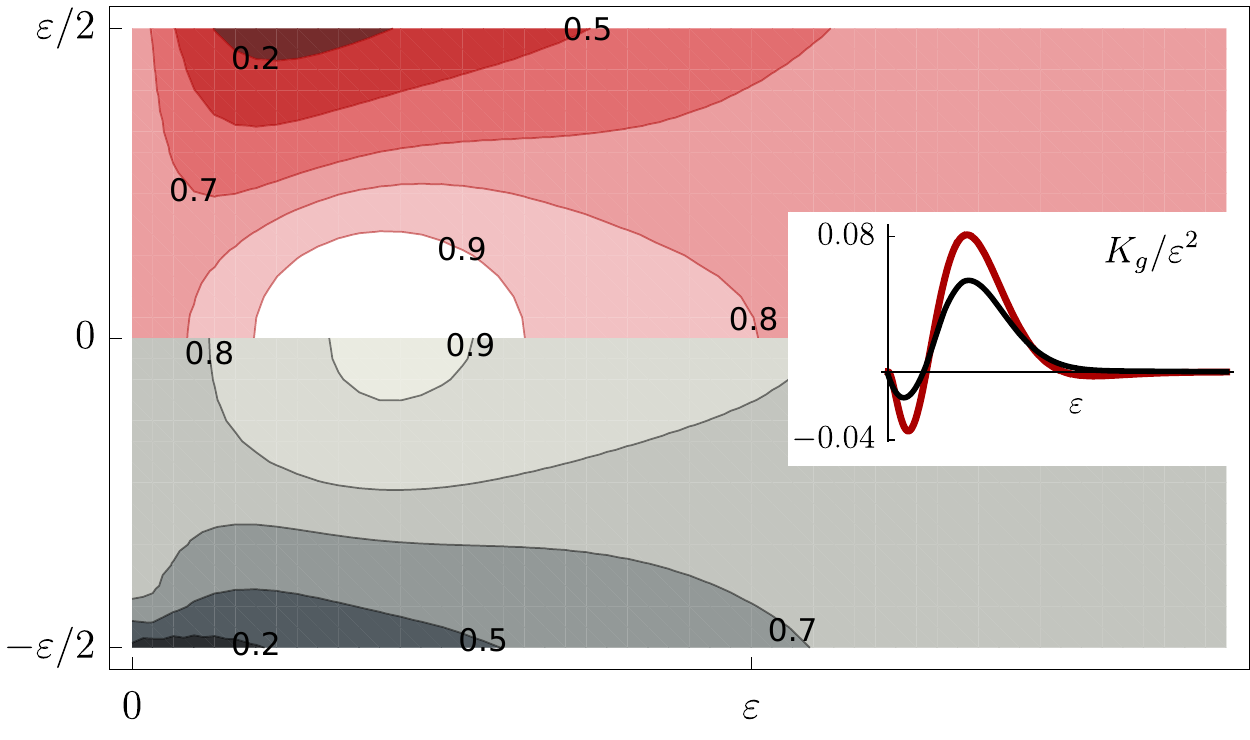}
		%	}
		%
		%		\subfloat[][]{
		%			\includegraphics[width=0.5\linewidth]{figures/likeaphoto}
		%			\label{likeaphoto}
		%		}
		%	\hfill
		\caption{Normalized level curves of the Gaussian curvature: the extensible rod predictions (upper part in red-tones) vs FvK FE predictions (lower part in gray-tones). The inset shows a comparison of the respective weighted averages along the axis. }
		\label{kgcomparison}
	\end{center}
\end{figure}

For the macroscopic behavior of the rod, the simple analytic expressions obtained by the inextensible model for $\chi^*$, $d^*$ and $v'(\ell)$ in section \ref{inextloc} seem surprisingly accurate, cfr. Figs.~\ref{fig:compchidstara}-\ref{fig:compchidstara}. However, the inextensible model, derived from the constraint for $\det \Kb$ to vanish almost everywhere in $\Omega$, could never describe neither the Gaussian curvature field nor the membrane stresses. To this aim, we compare the benchmark FvK results only with the extensible rod model. In Fig.~\ref{kgcomparison} the level curves of the Gaussian curvature are plotted. These level curves are rescaled to range within $0$ and $1$ corresponding respectively to the minimum and maximum values attained by both the models in $\Omega$: 
$$
0 \leftrightsquigarrow \min \{\min_{[0,\ell]} K_g^{\texttt{1d}},\min_\Omega K_g^{\texttt{FvK}}\} \quad \textrm{and}\quad 1 \leftrightsquigarrow \max \{\max_{[0,\ell]} K_g^{\texttt{1d}},\max_\Omega K_g^{\texttt{FvK}}\}.
$$ 
%and 
%$$
%1 \leftrightsquigarrow \max \{\max_{[0,\ell]} K_g^{\texttt{1d}},\max_\Omega K_g^{\texttt{FvK}}\}.
%$$
Being symmetric with respect to $y$, we have used the upper part to draw, in red tones, $K_g$ as predicted by the extensible rod and the lower part to draw in gray tones the results of a two-dimensional FE analysis with FvK. 
In the same Figure, the inset plots the weighted average of the two-dimensional field of Gaussian curvature and the reduced notion of Gaussian curvature, given by the right-hand side of \eqref{meandetK}. These results are in very good agreement with of the FvK model.

From the knowledge of the fields $v$  and $k$, Eq. \eqref{displ} allows to reconstruct the two-dimensional displacement, and then the deformed surface. 

\begin{figure}
	\begin{center}
		\small
		%	\hfill
		\subfloat[][]{
			\includegraphics[width=0.5\linewidth]{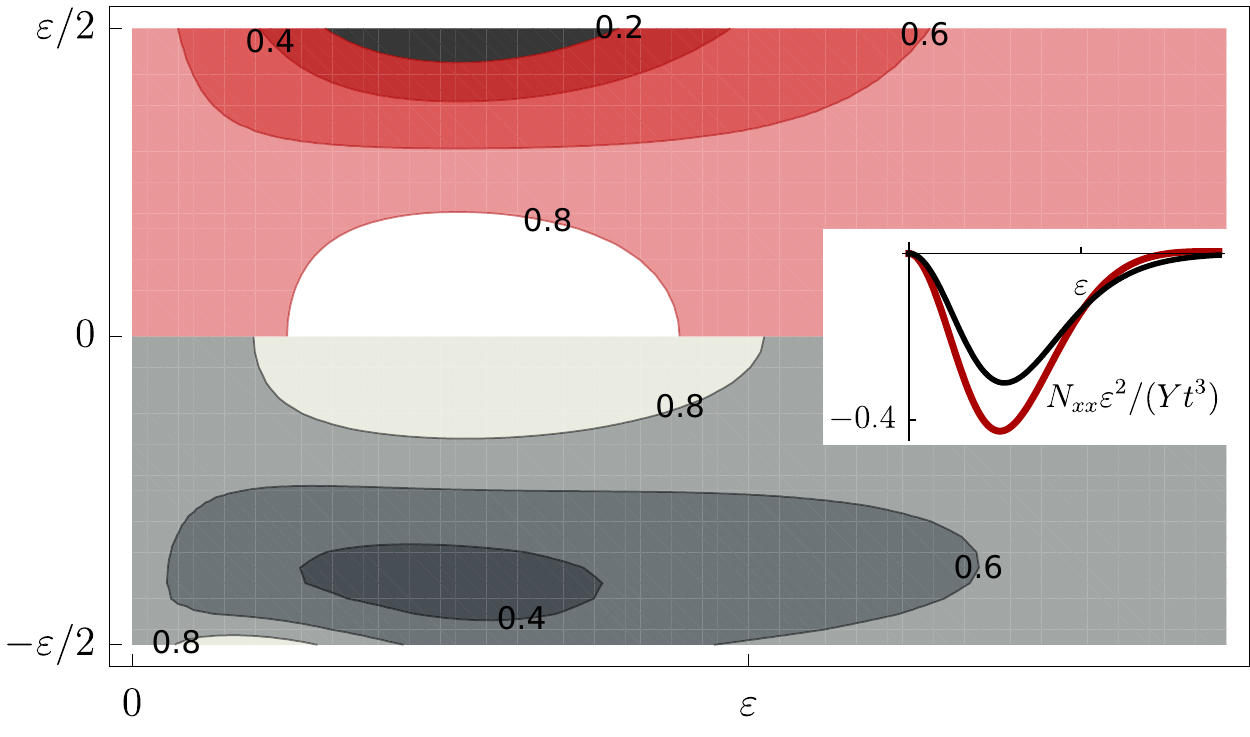}
			\label{nxx1d}
		}
		%
		%	\hfill
		\subfloat[][]{
			\includegraphics[width=0.5\linewidth]{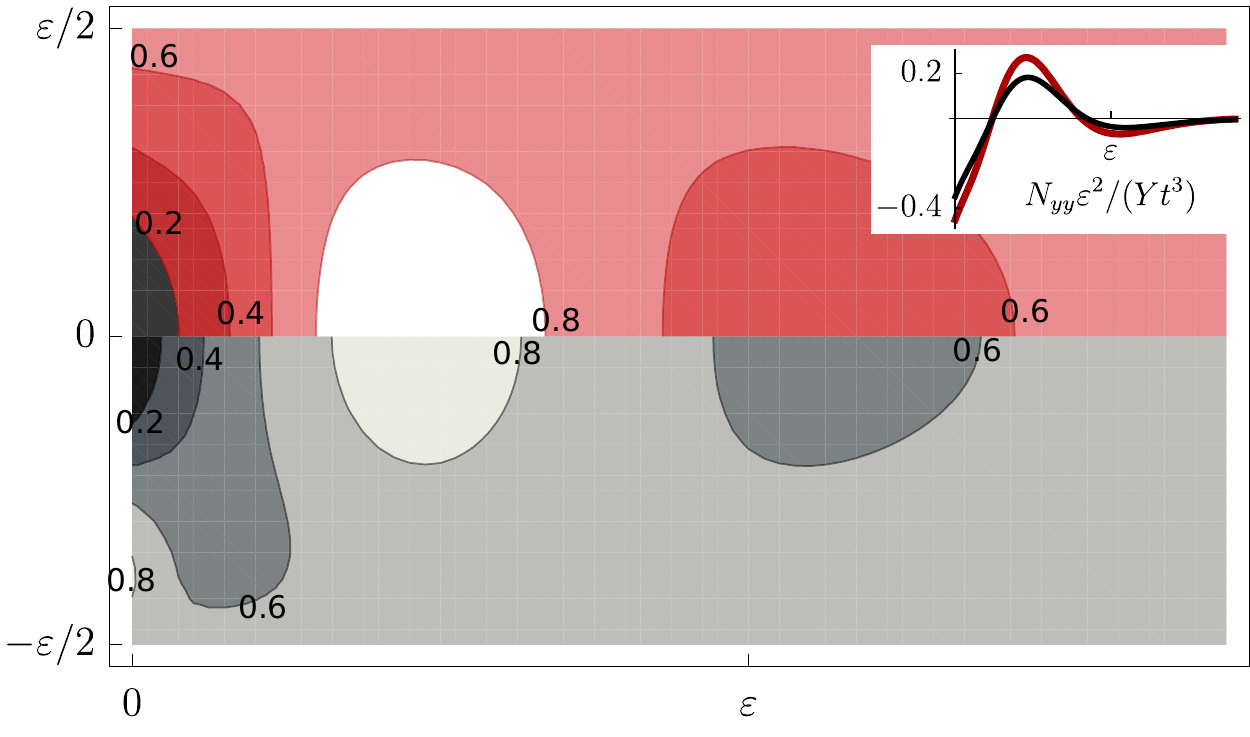}
			\label{nyy1d}
		}
		\hfill
		\caption{Normalized level curves for the membrane stresses $N_{xx}$ (a) and $N_{yy}$ (b): the extensible rod predictions (upper part in red-tones) vs FvK FE predictions (lower part in gray-tones). The insets show comparisons of the respective weighted averages along the axis.}
	\end{center}
\end{figure}

Finally, we remark that the extensible rod model allows for an estimate of the membrane stress fields via the scalar field $f$. Specifically, \tR{through} Eqns.~\eqref{displphi} and \eqref{Nij} we reconstruct the two-dimensional fields of the stress $N_{xx}$ and $N_{yy}$ and compare them to the ones of the FvK shell model in Figs.~\ref{nxx1d}-\ref{nyy1d}. 
Again a remarkable agreement is apparent also for the membrane fields. Slight discrepancies are localized at the edges $(x=0,y=\pm \varepsilon/2)$, \tR{where  our Ansatz  is not probably sufficient to catch the exact $y$-distribution of the stress fields and  more terms would be required}.

\section{Discussion and conclusions}\label{discuss}
We have presented two new models of thin-walled non-linear rods, whose main features are:
\begin{enumerate}
	\item The transversal section is not rigid. For this reason, an additional kinematical descriptor $k$ is introduced, accounting for the change in curvature  of the transversal section. The resulting 1D model is then endowed with a 1D notion of Gaussian curvature, keeping track of the 2D character  of the shell model from which we started.
	\item In standard dimensional reduction, starting from a two-dimensional model, the  effects of the compatibility kind of evaporate. In both our theories, a  compatibility condition coupling bending and membrane problems is deduced, endowing the model with a 2D character.
	\item Localization phenomena, such as d-cones, are captured by our models. Analytical estimates are possible in the inextensible case.
\end{enumerate} 

A key question may arise: is there any circumstance in which the role of the  1D compatibility equation \eqref{comp1d} (or \eqref{comp1din}   for the inextensible case) is particularly undeniable? 

If we confine the  attention to the inextensible model, satisfying the compatibility translates into requesting that the solution belongs to the cone  \eqref{cone}; the stress-free configuration (point $\Nb$ in Fig. \ref{conee}) belongs to $\Cc$; the  boundary conditions for $x=0$ compel the solution to tend to a point  belonging  to $\Cc$ as well; all in all, \textit{this particular set of  boundary conditions}  leads the solution to stay close to $\Cc$, even if no \textit{a priori} constraint is taken into account. 

One then might argue that the compatibility does not have a strong role, at least for this specific problem and minimizing the bending energy would suffice to obtain solutions sufficiently close to the cone. 
Nevertheless, regardless of the boundary conditions that activate or not the compatibility constraint,  the nonlocal (or inextensible) and the pure bending model differ for a crucial point: the bending energy is a positive quadratic functional, and then its direct minimization delivers a unique solution; this is not the case when the model is endowed with the  compatibility condition. Thus, multiple solutions are possible if the compatibility is taken into account, as we will see in the next subsection: besides localization phenomena, \textit{the 1D compatibility induces mutistability}.

\subsection{Compatibility and multistability: uniform curvature solution}
\label{sec:multistab}

%\begin{figure}
%	\centering
%	\includegraphics[width=0.5\linewidth]{figures/stability1.pdf}
%	\caption{Bistability regions (shaded) in the $(\alpha,\beta)$ plane for different Poisson ratios. Within the white region the uniform curvature configuration \eqref{unifconfig} is unstable; this includes isotropic materials where $\alpha=\beta=1$ (red point). The black point has been used to compare  FE simulations, based on FvK model, and analytical results; it  corresponds to $\alpha=11.52$, $\beta=1$, $\nu=0.851$.}
%	\label{bistability}
%\end{figure}

\begin{figure}
	\begin{center}
		\small
		%	\hfill
		\subfloat[][]{
			\includegraphics[width=0.45\linewidth]{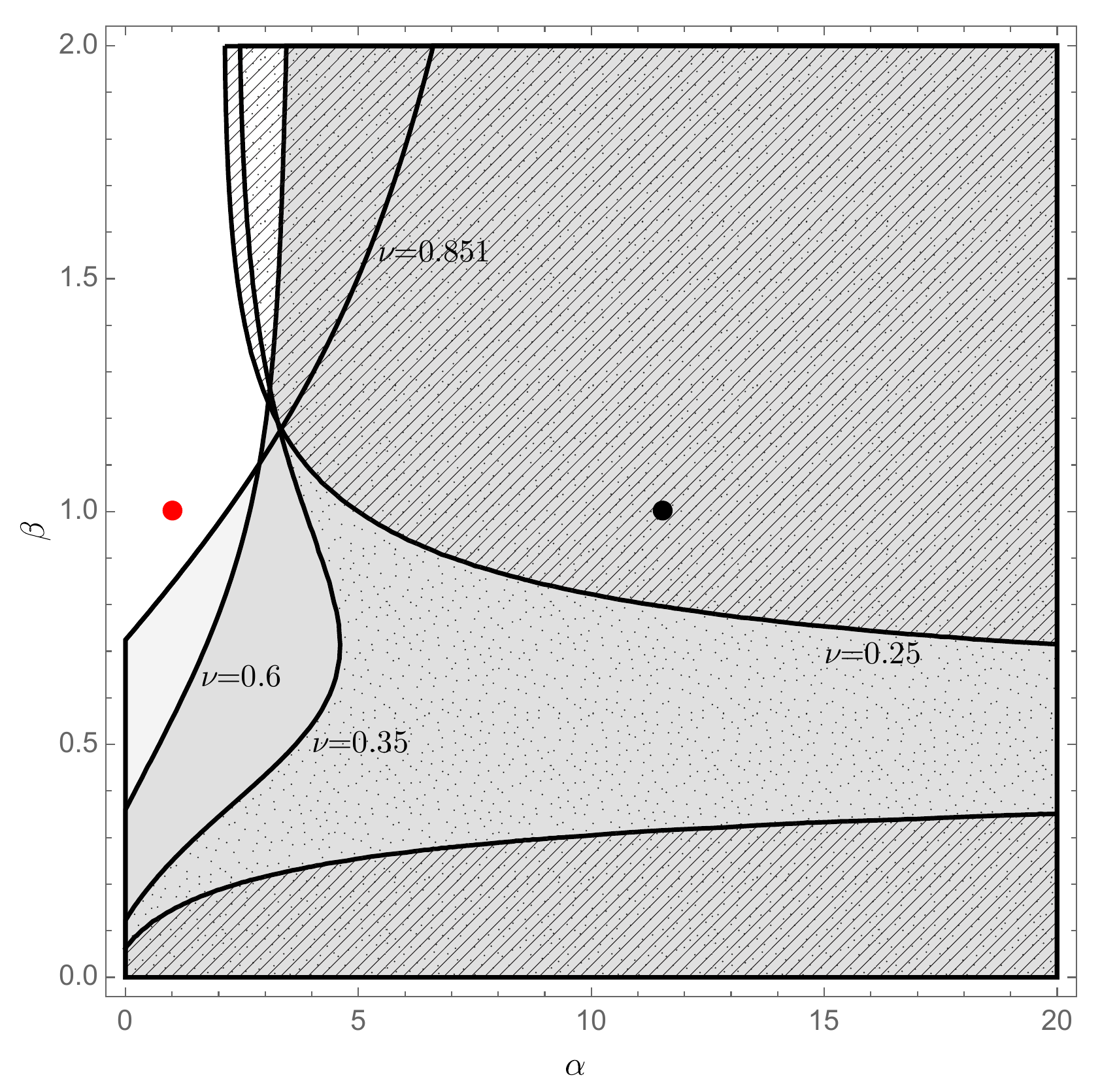}
			\label{bistability}
		}
		%
		%	\hfill
		\subfloat[][]{
			\includegraphics[width=0.45\linewidth]{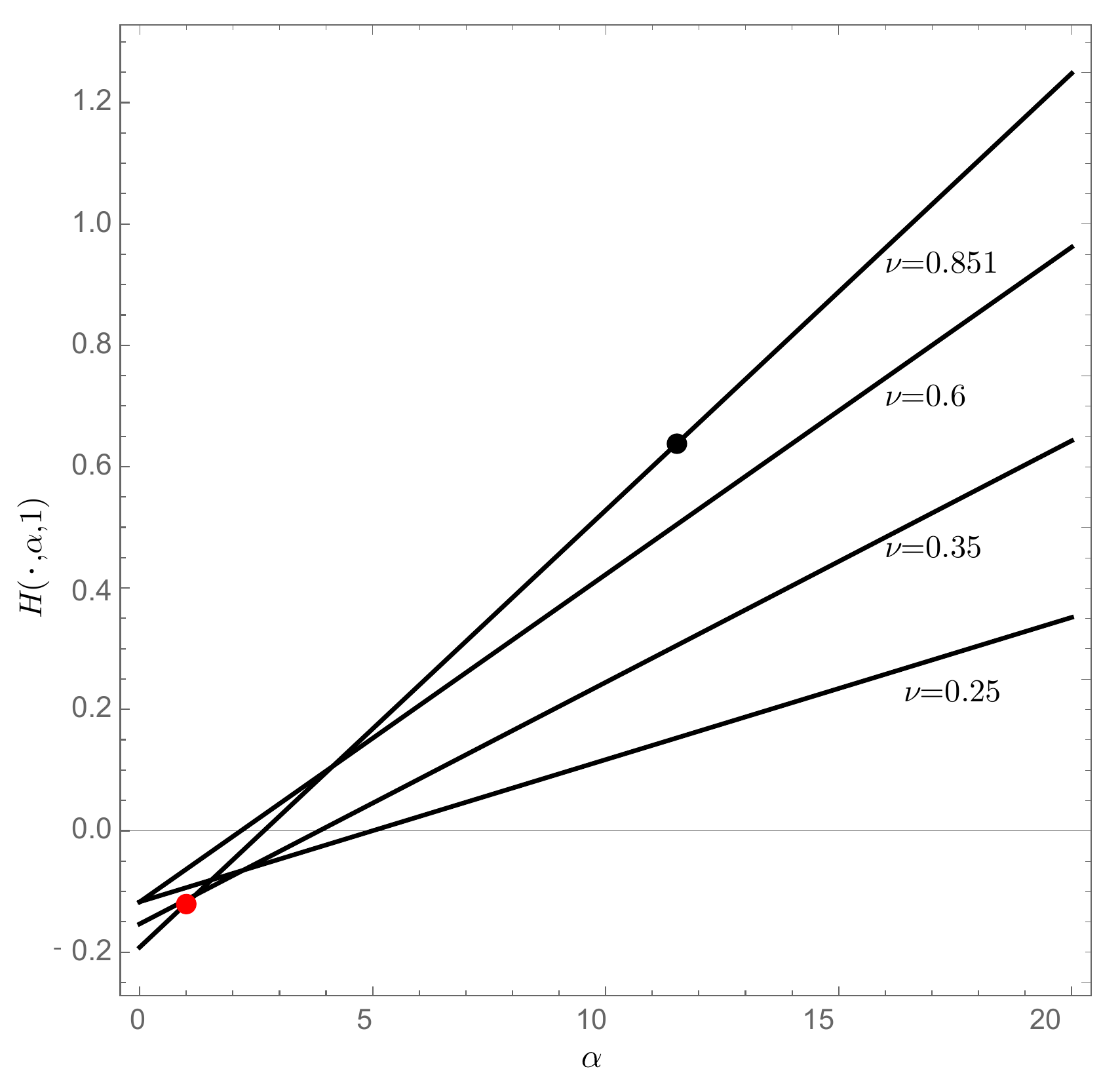}
			\label{bistabilitybeta}
		}
		\hfill
		\caption{Bistability regions (shaded) in the $(\alpha,\beta)$ plane for different Poisson ratios \tR{(a) and cross section of $H$ for $\beta=1$ and different Poisson ratios (b)}. Within the white region the uniform curvature configuration \eqref{unifconfig} is unstable; this includes isotropic materials where $\alpha=\beta=1$ (red point). The black point in  has been used to compare  FE simulations, based on FvK model, and analytical results; it  corresponds to $\alpha=11.52$, $\beta=1$, $\nu=0.851$.}
	\end{center}
\end{figure}

For the inextensible case the compatibility requires the condition $k \,v''=\varepsilon (k')^2/12$ to hold. This introduces a strong nonlinearity in the problem to solve which we introduced the cone coordinates $(c,\vartheta)$ in Sect.\ref{inextloc}. We show below that one can indeed have multiple equilibria and, in some cases, multiple stable equilibria.

The evaluation of the optimal $c$ in \eqref{cast} allows to obtain the bending energy on the cone $\mathcal{C}$ as a function of $\vartheta$
\begin{equation}\label{arricciata}
\begin{aligned}
\widehat{\Ec}_b(\vartheta)=&\widetilde{\Ec}_b\big(c^*(\vartheta),\vartheta\big)\propto\\
&\frac{k_0^2}{\varepsilon^2}\,\big( 3+\nu+(1-\nu)\cos\vartheta  \big)\,\sin^2\frac{\vartheta}{2}.
\end{aligned}
\end{equation}
This energy admits more than one stationarity point:  together with the solution presented in Sec. \ref{inextloc}, it is easy to see that $\widehat{\Ec}_b(\vartheta)$ has a stationarity point for $\vartheta=\pi$, a solution corresponding to 
\begin{equation}\label{unifconfig}
\chi(x)=v''(x)=\frac{\nu\,k_0}{\varepsilon}, \quad  k(x)=0, \quad \forall x\in [0,\ell].
\end{equation}
Indeed for $\vartheta=\pi$ we have $k'=0$ and therefore $k(x)=\text{const.}$; recalling the compatibility equation, \eqref{unifconfig} follows.  This configuration corresponds to the point blue in Fig. \ref{conee} (b). Hence, we could have a second possible configuration of the rod where all points have the same constant axial curvature, being the transversal curvature null.

The stability of this configuration can be studied on evaluating the second derivative of $\widehat{\Ec}_b(\vartheta)$ with respect to $\vartheta$ at the point $\vartheta=\pi$:
\begin{equation}
\partial^2_{\vartheta\vartheta}\widehat{\Ec}_b(\vartheta)\Big|_{\vartheta=\pi}=-\frac{1}{2}(1-\nu)\,\nu\,\frac{k_0^2}{\varepsilon^2},
\end{equation}
which is negative  for $0\leq\nu<1$. 

\begin{figure}[t]
	\centering
	\includegraphics[width=0.5\linewidth]{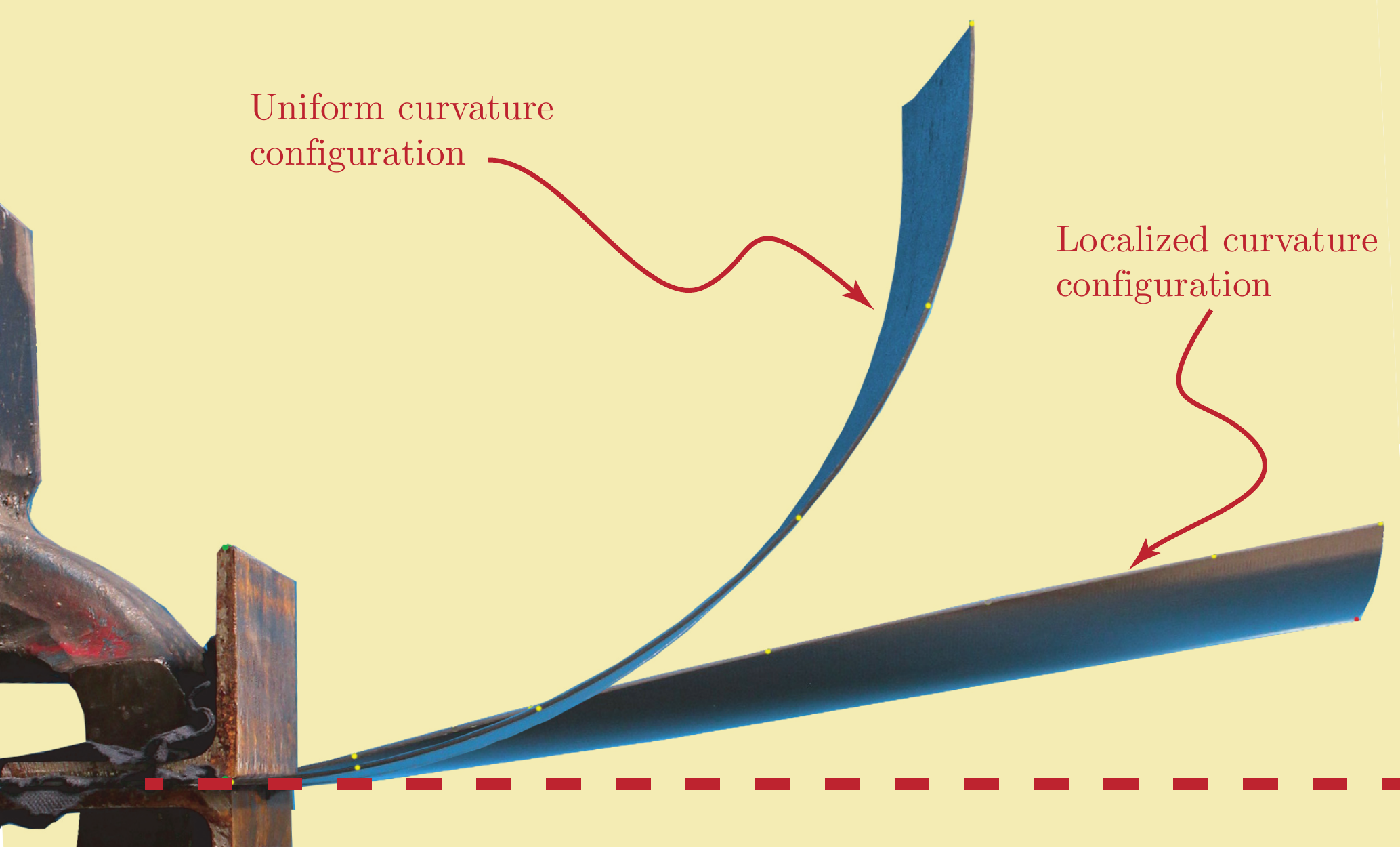}
	\caption{\tR{Experiment showing both stable configurations.}}
	\label{multiexper}
\end{figure}

Thus, the configuration \eqref{unifconfig} is not stable. \tR{We conclude that for an isotropic material, the only stable configuration is the localized one; however, removing the hypothesis of isotropy could lead to a different scenario.}
%However, the unstable character of this equilibrium  holds for isotropic materials: removing this last hypothesis could  lead to stability. 
To see this, let us consider an orthotropic material; the stiffness tensor $\Do$ is given by the following Voigt representation:
\begin{equation}
\Do=D\left(\begin{array}{ccc}
1& \nu & 0  \\ 
\nu & \beta & 0  \\ 
0 & 0 &\alpha\,(1-\nu)/2
\end{array} 
\right),
\end{equation} 
with $\alpha>0$, $\beta>0$, $-\sqrt{\beta}<\nu<\sqrt{\beta}$; the constant $\beta=E_2/E_1$ represents the ratio between the two Young moduli and $\alpha\,\frac{1-\nu}{2}$ represents the shear modulus. Isotropic materials are obtained when $\alpha=\beta=1$.

The change of coordinates \eqref{changecoord}, to diagonalize the bending energy, has to be replaced by
\begin{equation}\label{changecoordan}
\begin{aligned}
&\xi=\frac{1}{2}\sqrt{1+\frac{\nu}{\sqrt{\beta}}}\left(\sqrt{\beta}\,\frac{k}{\varepsilon}+ \chi\right), \quad \eta=\frac{1}{2}\sqrt{1-\frac{\nu}{\sqrt{\beta}}}\left(\sqrt{\beta}\,\frac{k}{\varepsilon}- \chi\right), \quad \zeta=\frac{\sqrt{1-\nu}\alpha}{2\sqrt{3}} (k')^2.
\end{aligned}
\end{equation}
The  cone \eqref{cone} of inextensible curvatures  then becomes:
\begin{equation}\label{conean}
\Cc:= \left\{ (\xi, \eta, \zeta) \;\Big|\; \frac{\eta^2}{c_1^2}+\frac{\zeta^2}{c_2^2}=\xi^2  \right\}, \qquad\quad c_1=\sqrt{\frac{\sqrt{\beta}-\nu}{\sqrt{\beta}+\nu}}, \quad c_2=\sqrt{\frac{\alpha\,(1-\nu)}{\sqrt{\beta}+\nu}}.
\end{equation}
%where
%\begin{equation}
%c_1=\sqrt{\frac{\sqrt{\beta}-\nu}{\sqrt{\beta}+\nu}}, \qquad c_2=\sqrt{\frac{\alpha\,(1-\nu)}{\sqrt{\beta}+\nu}}.
%\end{equation}
The cone $\mathcal{C}$ has then an elliptical cross section, whose semi-axes are in fact $c_1$ and $c_2$. The change of variable \eqref{change} then now reads:
\begin{equation}\label{change1}
\xi=c, \quad \eta=c\, c_1\cos\vartheta, \quad\zeta=c\,c_2\sin\vartheta,
\end{equation}
which allows to determine the analytical expression for the bending energy of the orthotropic rod $\widetilde{\Ec}_b(c,\vartheta)$. As in Section  \ref{inextloc}, we first evaluate the value $c^*(\vartheta)$ that makes stationary $\widetilde{\Ec}_b(c,\vartheta)$, and then deduce $\widehat{\Ec}_b(\vartheta)=\widetilde{\Ec}_b(c^*(\vartheta),\vartheta)$.
We do not the details here but, again, $\vartheta=\pi$ is a point that renders $\widehat{\Ec}_b(\vartheta)$ stationary.
Being $\partial^2_{cc}\tilde{\mathcal{E}}_b>0$ and $\partial^2_{c\vartheta}\tilde{\mathcal{E}}_b=0$, this stationary point is stable if the component of the Hessian 
\begin{equation}
\partial^2_{\vartheta\vartheta}\widehat{\Ec}_b(\vartheta)\Big|_{\vartheta=\pi}=H(\nu,\alpha,\beta),
\end{equation}
a function of the material parameters $\nu$, $\alpha$ and $\beta$, is positive. The analytical expression of $H(\nu,a,b)$ can be determined, but it is quite cumbersome and we do not report it. However, we plot, in Fig.~\ref{bistability} \tR{(a)}, the regions of the $(\alpha,\beta)$ plane where $H(\nu,\alpha,\beta)>0$ for several values of the Poisson \tR{ratio}. In  these shaded regions, there are at least two stable configurations: not only the localized-curvature solution discussed in the previous sections but also equilibrium $\vartheta=\pi$, corresponding to the configuration \eqref{unifconfig}, is stable. \tR{In Fig.~\ref{bistability} (b) we plotted the cross section of $H$ for $\beta=1$, which shows that there is not a sudden shift in the behavior, as those described within the framework of the catastrophe theory.}

For instance, when $k_0=6.67$ m$^{-1}$, $\ell=0.45$ m, $\varepsilon=\ell/3=0.15$ m, $h=1$ mm, $\nu=0.851$, $\alpha=11.52$ and $\beta=1$ (namely the black point in Fig.~\ref{bistability}), the inextensible rod model predicts by \eqref{unifconfig} $\chi(x)=5.67\,\text{m}^{-1}$, whilst the FvK FE computations find a very similar shape and predict an average axial curvature $\overline{K}_{xx}\simeq 5.35\text{m}^{-1}$. 
\tR{Both these configurations are shown in Fig. \ref{multiexper}.}

%\appendix
%\section{My Appendix}
%Appendix sections are coded under \verb+\appendix+.
%

%\bibliographystyle{cas-model2-names}
%\bibliography{bibtex}

\end{document}